\documentclass[12pt]{article}%\usepackage{cite}
\usepackage{fix-cm}
\usepackage{heck}
\usepackage{graphicx}
\usepackage{multicol}
\usepackage{amsmath}
\usepackage{amsfonts}
\usepackage{amssymb}
\usepackage{hyperref}
\usepackage{setspace}
\usepackage[all]{xy}

\def\spc{\hspace{.5pt}}
\usepackage{amssymb, amsmath, amsopn, amsthm}
\usepackage{psfrag}
\usepackage{subfigure}

\usepackage{color}
\usepackage{epsfig}
\usepackage{graphicx}
\usepackage{epstopdf}
\usepackage{mathtools}
\usepackage{bbold}
\def\be{\begin{equation}}
\def\ee{\end{equation}}

\begin{document}

\setlength{\textwidth}{16.2cm}
\setlength{\textheight}{21.8cm}
\addtolength{\oddsidemargin}{0mm}
\addtolength{\topmargin}{-10mm}

\date{October 2012}
%\preprint{UCD-HEP-???}
\title{\LARGE\bf  Black Hole Entanglement
  \\[-7mm] { and} \\[2mm]
Quantum Error Correction}

\institution{UvA}{\centerline{${}^{1}$Institute for Theoretical Physics, University of Amsterdam, Amsterdam, The Netherlands}}

\institution{PU}{\centerline{${}^{2}$Department of Physics, Princeton University, Princeton, NJ 08544, USA}}

\institution{PCTS}{\centerline{${}^{2}$Princeton Center for Theoretical Science,  Princeton, NJ 08544, USA }}

\authors{Erik Verlinde\worksat{\UvA}\footnote{e-mail: {\tt E.P.Verlinde@uva.nl}} and Herman Verlinde\worksat{\PU,\PCTS}\footnote{e-mail: {\tt verlinde@princeton.edu}}}

\def\spc{\hspace{.5pt}}

%\date{November 19, 2007}

\abstract{It was recently argued in \cite{amps} that black hole complementarity strains the basic rules of quantum information theory, such as monogamy of entanglement.
Motivated by this argument, we develop a practical framework for describing black hole evaporation via unitary time evolution,
based on a holographic perspective in which all black hole degrees of freedom live on the stretched horizon. We model the horizon as a unitary
quantum system with finite entropy, and {\it do not postulate  that the horizon geometry is smooth}. We then show that, with mild assumptions, one can reconstruct local effective
field theory observables that probe the black hole interior, and relative to which the state near the horizon looks like a local Minkowski vacuum.
The reconstruction makes use of the formalism
of quantum error correcting codes, and works for black hole states whose entanglement entropy does not yet saturate the Bekenstein-Hawking bound.
Our general framework clarifies the black hole final state proposal, and allows a quantitative study of the transition into the
``firewall'' regime of  maximally mixed black hole states.}

\addtolength{\abovedisplayskip}{1mm}
\addtolength{\belowdisplayskip}{1mm}

%\keywords{}

\maketitle
\def\mathbi#1{\textbf{\em #1}} 
\def\som{{ \textit{\textbf s}}} 
\def\tom{{ \textit{\textbf t}}} 
\def\nom{{ \textit{\textbf n}}} %{\mbox{\fontsize{13pt}{.15pt}${ \smpc \mathbf{n}\smpc }$}}}
\def\mom{{ \textit{\textbf m}}} %{\mbox{\fontsize{13.5pt}{.15pt}$\mathbf m$}}}
\def\kom{{ \textit{\textbf k}}}  %\mbox{\fontsize{13.5pt}{.15pt}$\mathbf m$}}}
\def\nomt{{ \textit{\textbf n}}}  %\mbox{\fontsize{9.5pt}{.1pt}\smpc $\mathbf n$}}}
\def\momt{{ \textit{\textbf m}}}  %\mbox{\fontsize{9.5pt}{.1pt}$\mathbf m$}}}
\def\komt{{ \textit{\textbf k}}} %mbox{\fontsize{9.5pt}{.1pt}$\mathbf m$}}}
%\def\kom{{\textbf{\em k}}}
%_{{}%_{\mbox{\fontsize{4pt}{.5pt}{\omega}}}}}
%\def\mom{{\textbf{\em m}}}
%_{{}_{\mathtt{\omega'}}}}}
\def\la{\langle}
\def\bea{\begin{eqnarray}}
\def\eea{\end{eqnarray}}
\def\is{\! & \! = \! & \!}
\def\ra{\rangle}
\def\half{{\textstyle{\frac 12}}}
\def\cL{{\cal L}}
\def\bigll{\bigl}
\def\bigrr{\bigr}
\def\halfi{{\textstyle{\frac i 2}}}

\def\ba{\begin{eqnarray}}
\def\ea{\end{eqnarray}}
%\section{ \label{sec:}}

%\section{}

%\subsubsection{}

%\section{Introduction}

%\subsubsection{Experimental results}

\def\ibar{{\, {\overline{i}}\, }}
\def\kbar{{\spc {\bar{k}\spc }}}

\newcommand{\rep}[1]{\mathbf{#1}}
\newcommand{\Tr}{\, {\rm Tr}}
\def\betaH{\beta}
\def\XX{\text{X}}
\def\IR{\text{IR}}
\def\UV{\text{UV}}
\def\GUT{\text{GUT}}
\def\singlet{\text{GUT}}
\def\be{\bea}
\def\ee{\eea}
\def\delbar{\overline{\partial}}
\newcommand{\smpc}{\hspace{.5pt}}
\def\ra{\bigr\rangle}
\def\la{\bigl\langle}
\def\ccdot{\!\spc\cdot\!\spc}
\def\AA{{\raisebox{-1pt}{\scriptsize \nspc $A$}}}

\def\BB{{\raisebox{-1pt}{\scriptsize\nspc $B$}}}
\def\BH{{\raisebox{-1pt}{\scriptsize \nspc $B\nspc H$}}}
\def\R{{\raisebox{-1pt}{\scriptsize \nspc $R$}}}
%{{B\nspc H}} %{{}_{\!\mbox{\fontsize{7pt}{.5pt}{$B\nspc H$}}}}}%{{}_{\! B\! \smpc H}}}
\def\nspc{\!\spc\smpc}
\def\uU{\mbox{\textit{\textbf{U}}}}
\def\cC{\mbox{\textit{\textbf{C}\!\,}}}
\def\bn{\mbox{\textit{\textbf{n}\!\,}}}
\def\bbn{\mbox{\scriptsize{\textit{\textbf{n}}}}}
\def\pP{\mbox{\textit{\textbf{P}\!\,}}}
\def\rR{{\textit{\textbf{R}\!\,}}}

\def\sS{\uU}
\def\iI{\mbox{\textit{\textbf{I}\!\,}}}
\def\hH{H} %\mbox{\textit{\textbf{H}\!\,}}}
\def\vV{\mbox{\textit{\textbf{V}\!\,}}}
%\def\bbB{{{}_{\! \mbox{\fontsize{5pt}{.5pt}{$B$}}}}}
%\mbox{\scriptsize \tiny $B$}}}}
\def\aaA{{{}_{\! \mbox{\scriptsize  $A$}}}}

\def\bBB{{{}_{\! \mbox{\scriptsize  $B$}}}}
\def\ccC{{{}_{\nspc \mbox{\scriptsize  $C$}}}}
\def\abAB{{{}_{\! \mbox{\scriptsize  $A\nspc\nspc B$}}}}
\def\bB{{\mbox{\scriptsize $b$}}}
\def\bbB{{\mbox{\scriptsize  $b$}}}
\def\eE{{}} %{\mbox{\fontsize{7pt}{.5pt}{$R$}}}}%\mbox{\tiny \it BH}}}}
\def\aA{\mbox{\textit{\textbf{A}\!\,}}}
\def\pPi{\mbox{\textit{\textbf{$\Pi$}\!\,}}}
\def\Error{\mbox{\textit{\textbf{E}\!\,}}}
\def\QC{{{}_{\!\! Q\! \smpc C}}}
\def\xX{{\mbox{\footnotesize X}}}
\def\im{{\rm i}}
\def\ccc{{{\!}_{\mbox{\small {$c$}}}}}
\def\aaa{{{\nspc}_{\mbox{\footnotesize {$a$}}}}}
\def\aaaa{{{\mbox{\fontsize{8.5pt}{.11pt}{$a$}}}}}
\def\bbb{{{\nspc}_{\mbox{\footnotesize {$b$}}}}}
\def\bbbb{{{\mbox{\fontsize{8.5pt}{.11pt}{$b$}}}}}
\def\pp{{\mbox{\fontsize{8.5pt}{.11pt}$\mathbf\sf W$\!}}}
\def\tr{{\rm tr}}
\def\oneoverN{{1\over N}} %\frac{\raisebox{-1pt}{\small $1$}}{\raisebox{.5pt}{\small $N$\spc }}}

\def\oneovertN{{1\over N}} %\frac{\raisebox{-1pt}{\small $1$}}{\raisebox{.5pt}{\small ${2N}$\spc }}}
\def\oneoversqN{{1 \over \sqrt{N} }} %\frac{\raisebox{-1pt}{\small $1$}}{\raisebox{.5pt}{\small $\sqrt{N}$\spc }}}
\def\oneoverZ{\frac{\raisebox{-1pt}{\small $1$}}{\raisebox{.5pt}{\small $Z$\spc}}}
\def\oneoverNZ{\frac{\raisebox{-1pt}{\small $1$}}{\raisebox{.5pt}{\small $N\nspc Z$\spc}}}
\def\iii{i}
\def\jjj{j}
\def\bh{{\mbox{\fontsize{7pt}{.7pt}{$BH$}}}}%\mbox{\tiny \it BH}}}}
\enlargethispage{\baselineskip}

\setcounter{tocdepth}{2}
\tableofcontents

\newpage
\addtolength{\baselineskip}{.25mm}
\addtolength{\parskip}{.3mm}
\renewcommand\Large{\fontsize{15.5}{16}\selectfont}

\section{Introduction}

The question of how black holes absorb, store, share and release quantum information remains among the most vexing mysteries
in theoretical physics \cite{hawking,unruh}.  Logical consistency of quantum black hole physics with the Bekenstein-Hawking entropy  
\cite{bekenstein,hawking} imposes tight constraints on the microscopic theory. The discovery of the holographic principle  
\cite{hoofthologram}\cite{susskindhologram} and its realization in string theory \cite{maldacena} suggests that  black holes satisfy all rules of 
quantum mechanics.   The principle of black hole complementarity \cite{susskindcompl} is designed to reconcile the seemingly 
conflicting observations of  an outside observer and an infalling observer. It postulates that\\[2.5mm]
${}$~~\parbox{15.5cm}{\addtolength{\baselineskip}{1mm}
i)\;\; black hole formation and evaporation are described via unitary quantum evolution 
ii)\; the region outside the stretched horizon is well described by QFT in curved space
iii) to an outside observer, the stretched horizon behaves like a quantum mechanical  ${}$~~~~${}$\,\spc membrane
with microscopic entropy bounded by the Bekenstein-Hawking formula iv) an infalling observer can cross the event horizon without encountering any trouble.}

\medskip

In a recent paper \cite{amps}, Almheiri et al formulated an interesting counter-argument against the simultaneous validity of these four assumptions. 
The reasoning of \cite{amps}, referred to as the ``firewall argument",  goes as follows. At late times the quantum state of the black hole is maximally entangled with the early radiation. Because entanglement can not be shared, and the information eventually has to come out, the late radiation can not be entangled with the black hole. Since the latter property is believed to be a necessary requirement for the quantum state to look like the local Minkowski vacuum near the horizon, one seems forced to conclude that the infalling observer can not pass through the horizon without experiencing dramatic events. This provocative conclusion has led to a lively debate \cite{followup}\cite{susskind-entanglement}.

The firewall argument is the old information paradox in reverse. The original paradox was that 
locality and the smoothness of the horizon seems to imply that quantum mechanics falls short in describing black hole formation and evaporation.
So if instead one adopts the rules of quantum mechanics and quantum information theory, smoothness of the black hole horizon becomes the real mystery \cite{mathur-info-paradox}.
The value added by the reasoning of \cite{amps}, and of other recent  \cite{haydenpreskill, fastscrambler, avery, giddings-qu-info} and less recent  \cite{page-subsystem, page-bh-info} works, is that it aims to bring the paradox more sharply into focus, by adopting a
more precise  quantum information theoretic language. 

The ramifications of the firewall argument are somewhat disturbing, since it could endanger the validity of HawkingÕs derivation of the evaporation process,
which stands at the very basis of this scientific discussion. Therefore, before accepting its conclusion, one should carefully
 examine the logic and assumptions that go into it. This requires a precise enough framework, that preserves all the essential physics and elements of the paradox.

Motivated by this challenge, we develop a systematic description of the evaporation process that
keeps the rules of quantum mechanics manifest.
The idea is to treat the black hole as an ordinary quantum system that interacts with a radiation field through 
absorption and emission, much like an atom would. We  will {\it not adopt postulate} iv), that the horizon is smooth, but instead we will investigate
if we can {\it derive postulate} iv) from the other three postulates. The argument of \cite{amps} indicates that this derivation
may be possible for young black holes, that do not yet saturate the Bekenstein-Hawking bound, but should break down for old black holes that do saturate the bound.

Besides the postulates i), ii) and iii), our approach relies on one additional physical assumption: we assume that
 the quantum mechanical amplitudes, that describe the transition between different black hole states as a result of the emission of Hawking quanta,
 are ergodic matrices \cite{fastscrambler}. Their detailed form depends on unknown Planck  scale physics,  but their coarse grained properties are determined via standard
 thermodynamic reasoning. 
 
 As another helpful tool,  we will make use of the parallel, already exploited in \cite{haydenpreskill}, between the physics of black hole radiation
and  the loss of coherence of a quantum computer, due to its interaction with a noisy environment. 
This analogy allows one to apply inventive
theoretical techniques of quantum information theory \cite{quantuminformation, preskilllectures}, such as quantum error correcting codes (QECC's), to organize
the internal state of the black hole \cite{haydenpreskill}.
QECC's are designed to counteract  the quantum information loss.  We will show that in the black hole context,  the same technique
can be used to safeguard the smoothness of the horizon. It enables the construction of microscopic observables, that play the role of Hawking partners
of the outside radiation, relative to which the black hole state looks like a local horizon vacuum state. 
Smoothness of the horizon and the Hawking evaporation process are thus consistent with each other, and with the laws of quantum mechanics --
 at least for black hole states that do not saturate the BH entropy.

QEC codes are not full proof \cite{preskilllectures}. 
As long as the black hole state has a less-than-maximal entropy, the code works with high fidelity:
the local horizon geometry is smooth, and effective QFT is valid on both sides. When the black hole starts saturating the BH entropy bound, however,
the QEC procedure becomes less reliable and sometimes fails  -- this are the instances when quantum information 
is leaking out \cite{haydenpreskill}. Since the code is also responsible for the reconstruction of 
semi-classical interior observables,  the effective QFT description of the black hole interior breaks down in this limit.
 
In section 2, we state our man assumptions and develop an interaction picture  for black hole evaporation.
In section 3, we present the reconstruction of the black hole interior via QEC technology, and analyze the firewall limit.
For the well-informed reader that prefers the short version of our story, we have summarized our main arguments in section 4.

\section{Setting the Stage}
\vspace{-2mm}

In this section and the next, we will describe the basic setup and main assumptions. Our guiding principle is that a black hole is an ordinary quantum mechanical system, 
-- much like an atom, albeit a very large one and with an enormous level density of states -- that evolves and interacts with its environment via standard hamiltonian evolution.  We start by setting up some notation that will 
help us model the evaporation process in a way that keeps the rules of quantum mechanics manifestly intact, in particular it obeys unitarity. Along the way, we present a version of the firewal argument, that can be directly addressed and tested within our framework.

\subsection{Black Holes States: Old and Young}

Consider a macrosopic black hole.
Suppose that we have measured its mass $M$ with some high accuracy.  For now, we will ignore the uncertainty in $M$, and any
other macroscopic properties such as angular momentum and charge.
We associate a Hilbert space ${\cal H}_{{}_{\! M}}$ to the black hole with mass $M$, whose states
\be
\bigll|\,\iii\,\bigrr\rangle\in\; {\cal H}_M
\ee
are indistinguishable from a macroscopic perspective. Each state $\bigl| \, i \, \ra$ describes the interior of a black hole, with microscopic properties that are hidden from the outside observer by the stretched horizon. The number of black holes micro-states is determined by the Bekenstein Hawking entropy\footnote{We assume here that we have performed a suitable coarse graining, since strictly speaking the number of quantum states of a black hole should be represented in terms a density of states. This technical point will not be important for our subsequent discussion.} 
\be
N \, = \, {\rm dim}\, {\cal H}_M \, =\,  e^{S_{BH}},
\ee
with $S_\BH = 4\pi M^2$. 
We will assume that the time evolution of every micro state describes a black hole that slowly evaporates via the Hawking process. So as time progresses, every state $\bigll|\, i\,\bigrr\rangle \in {\cal H}_M$ will evolve into a new state  given by the product of a micro-state of a black hole with smaller mass $M' = M-E$ and a radiation state with energy $E$.  

To describe this evaporation process we need to enlarge our Hilbert space, so that it incorporates all possible intermediate situations. So we introduce the total Hilbert space $\cal H$, given by the tensor product
\be
\label{hilbtotal}
{\cal H} = {\cal H}_\BH \otimes {\cal H}_R,
\ee
where ${\cal H}_\BH$ is the space spanned by all black hole states with mass less than $M$
\be
\label{bhilb}
 {\cal H}_\BH \! =\ \oplus_{\strut \!\!\!\!\!\!\!\!\!\!{}_{E\leq M}}{\cal H}_{M-E},
\ee 
and  ${\cal H}_R$ is the Hilbert space of the radiation field outside of the stretched horizon. 
 In the next subsection, we will set up a general description of the
evaporation process as a unitary Hamiltonian evolution on the total Hilbert space (\ref{hilbtotal}). 
Note that the black hole states $\bigll|\, i\,\bigrr\rangle$ 
are not part of the asymptotic past and future Hilbert space, but are to be regarded as very long lived resonances.  It is standard in quantum mechanics to include resonances as an additional
 Hilbert space sector, even if they do not appear as true stable asymptotic states. 
 Unitarity is preserved under time evolution, because the resonant black hole states eventually all evolve
into the unique zero mass ground state $\bigl| 0 \ra_\BH$ times an asymptotic radiation state.
 We first summarize what one would expect the time evolution to look 
like from a somewhat coarse grained perspective.

First consider a young black hole of mass $M$. Let us imagine that it was formed in a pure state $\bigll |\, i\, \ra$. The total state of the black hole
with its environment then takes the form
\be
\label{initialy}
\bigll|\Psi_{\rm young} \bigrr\rangle \, =\, \bigll|\spc \, \iii\, \bigrr\rangle
\,\bigll|\spc   \Phi\bigrr \rangle_\eE .
\ee
Here $\bigl|\Phi\ra$ denotes the state of the environment. Since the black hole is assumed to be formed in a pure state, there is no entanglement 
between the black hole and its environment.
Tracing over the environment yields the pure state density matrix 
\be
\label{young}
\rho^{\spc \rm young}_\BH \,
=  \bigll |\, i\, \ra \la \,i \,\bigrr|.
\ee
Next consider an old black hole of mass $M$ that has formed in the far past and has evaporated for a very long time. At this late stage, the black hole and radiation are described by a very-close-to-maximally entangled quantum state of the form
\be
\label{initial}
\bigll|\Psi_{\rm old} \bigrr\rangle\, =\, \frac{1}{\sqrt{N}} 
\sum_{\iii} \ \bigll|\spc \, \iii\, \bigrr\rangle
\,\bigll|\spc   \Phi_\iii \bigrr \rangle_\eE .
\ee
Here the sum runs over all basis states $\bigl|\,\iii\,\ra$ of the Hilbert space ${\cal H}_{M}$. 
The precise form of the states $\bigl|\Phi_i\rangle \in {\cal H}_R$ depends on the early history of the black hole. 
Since the dimension of ${\cal H}_{R}$ is much larger than that of ${\cal H}_{M}$, the states  $\bigl|\Phi_i\rangle$ can be safely assumed to form an orthonormal set. Hence when we trace the density matrix $\rho = |\Psi\ra \la  \Psi|$ over the radiation Hilbert space ${\cal H}_{R}$, the resulting black hole density matrix $\rho_\BH$  becomes, to a very good approximation,  proportional to the identity matrix on the Hilbert space ${\cal H}_{M}$ 
\be
\label{maxmix}
\rho^{\spc \rm old}_\BH \,
= \, \oneoverN %\frac{1}{N} 
\sum_{\iii}\,\bigll |\,\iii\, \ra \la \,\iii\,\bigrr|
\equiv\; \oneoverN \, \mathbb{1}_{M} .
\ee
In the following, we will consider the quantum states of both old as well as young black holes and take them as our initial condition at some time $t=0$. 
Since hamiltonian evolution is a linear process one can express the density matrix of an entangled black hole at time $t$ in terms of the evolved quantum states of initially unentangled black holes. We will make use of this fact to study the evolution of the density matrix $\rho_{BH}$ as the black hole continues to evaporate, and to see if it is possible to uncover an effective description of the semi-classical vacuum state near the horizon, as seen by an infalling observer. Note that even for young black holes this is not automatic, since we have not assumed the existence of an interior part. One of the main results of our paper is that this can indeed be done, with only a minimal set of assumptions. 

The maximally entangled old black holes and completely unentangled young black holes are the extreme limits of a broad class of intermediate cases going from slightly to almost maximally entangled situations. We will also consider these intermediate cases, since they are more realistic.  Indeed one could argue that even young black holes
can never be close to a pure state. As emphasized e.g. in \cite{vanraamsdonk},  in order for two different space time regions to be connected parts of the same space time, these regions have in 
terms of the microscopic physics to be maximally entangled.  So in this sense, even young black holes may already come close to saturating the BH entropy bound.

\def\mn{{{}_{\raisebox{-.5pt}{\scriptsize $\mom \nom$}}}}
\def\nm{{{}_{\raisebox{-.5pt}{\scriptsize $\nom \mom$}}}}

\subsection{The Firewall Argument}

We now briefly summarize one version of the argument in \cite{amps}. Consider the old black hole.
Following \cite{amps}, we factorize the total Hilbert space into an early and late part  
\be
{\cal H} = {\cal H}^{\rm \, late} \otimes {\cal H}_R^{\rm \, early}.
\ee
Here ${\cal H}_R^{\rm \, early}$ contains all radiation emitted during the early life of the black hole before $t=0$. 
The late Hilbert space ${\cal H}^{\rm late}$ decomposes into the product of the
Hilbert space of black hole states with mass $M'<M$, times the Hilbert space of all radiation that will be emitted at late times $t>0$. 
For notational convenience we will denote the two tensor factors of ${\cal H}^{\rm late}$ by
\be
{\cal H}^{\rm late} = {\cal H}_\AA \otimes {\cal H}_\BB, \qquad \quad 
{\cal H}_\AA = {\cal H}_\BH,  \qquad \quad  {\cal H}_\BB = {\cal H}_R^{\rm late}.
\ee
Indeed, the firewall argument involves an Alice and Bob gedanken experiment. Alice passes through the horizon at some time $t=\tau >0$ and observes
the interior black hole state in ${\cal H}_\AA$.  Bob, on the other hand, initially stays outside and then falls through the horizon a small time later.
He then catches up with Alice to confirm that she safely fell through the horizon. 
Suppose he succeeds and concludes that both are unharmed, for now. Why is this a problem?
To explain the issue, we need a bit more notation. 

At $t=\tau$,  the moment that Alice falls in, the black hole state (\ref{maxmix})
has evolved into a mixed state on the tensor product Hilbert space ${\cal H}^{\rm late}$ and also contains some radiation. 
Let us denote the orthonormal basis of late radiation states by
$$
\bigll| \,\nom\,\bigrr \rangle\, \in {\cal H}_\BB.
$$
The states $\bigll| \,\nom\,\bigrr \rangle$ span the Hilbert space of a quantum field theory in the curved space-time background in the neighborhood
of the black hole. In what follows, we will not need any detailed information about this QFT: all our arguments and calculations will go through 
regardless of whether it is strongly interacting or free. But for concreteness,  consider the case of a free field theory. We can then 
expand the fields as usual in terms of creation and annihilation modes via 
$\phi(t) = \sum_\omega {1\over \sqrt{\omega}} \left( b^\dagger_\omega e^{i\omega t} + b_\omega e^{-i\omega t}\right).$
Our short hand notation $\bigll| \,\nom\,\bigrr \rangle{}$  is then short-hand for the usual particle number eigenstates
\be
\label{fockstates}
\bigll| \,\nom\,\bigrr \rangle_\bB\, =
\,\prod_{\omega}
{1\over \sqrt{n_\omega!}} 
\bigll( b^\dagger_\omega\bigrr)^{n_\omega}\, \bigll| \,0 \,\bigrr \rangle_\bB.
\ee
In particular, the states $\bigl| \spc\nom\spc\bigr\rangle_\bB$ are energy eigenstates of the Schwarzschild Hamiltonian $H$.

The central player in the firewall argument, and our main object of study, is the density matrix of the black hole and late radiation at time $t= \tau> 0$. It is defined on
the tensor product Hilbert space ${\cal H}_\AA\otimes {\cal H}_\BB$ and takes the general form
\bea
\label{rhonm}
\rho_{\abAB} = \sum_{\nom, \mom}\spc  \rho_\nm \, \bigl|\spc \nom\spc \ra_\bB \la\spc \mom\spc \bigr|  .
\eea
The partial density matrix $\rho_\nm$  acts only on ${\cal H}_\AA$,
and specifies the correlation between the black hole state and external radiation.  
We do not have much a priori information about its detailed form, although in the following we will find out
quite a bit about its coarse grained properties. Indeed, we have one important piece of knowledge:
thanks to Hawking, or more directly (given that we already postulated the size of the BH Hilbert space) by application of standard rules of  statistical physics, we know that if we
take the partial trace of $\rho_\abAB$ over the internal Hilbert space, $\rho_\BB =  \tr_\AA\bigl(\spc \rho_\abAB\spc\bigr) $,
we obtain a thermal density matrix  
\be
\label{thermal}
\qquad \rho_\BB  =   
\sum_\nom \pp_\nom\, \bigll|\, \nom \, \ra_\bB \la\, \nom \, \bigrr| \qquad ; \qquad \pp_\nom = \frac{e^{-\betaH E_\nom}}{ Z} ,
\ee
with $Z\! =\! \sum_\nom\!  e^{-\beta E_\nom}$ and $\beta = 8\pi M$ the inverse Hawking temperature. Here we have normalized the Boltzman weights $\pp_n$ such that 
\be
\tr_\BB \bigr( \rho_\BB\bigr) = \sum_\nom \pp_\nom = 1.
\ee 
Note that equation (\ref{thermal}), although it looks static, in fact represents a time dependent state, by virtue of the fact that the level density of allowed frequencies $\omega$
increases inversely proportional with the time interval between $0$ and  $\tau$. So (\ref{thermal}) describes a radiation state with total 
average energy $\bar{E} = \sum_\nom E_n \pp_n$ that increases linearly with time $\tau$. In the following we will assume that the time $\tau$ is short compared to
the black hole life time, so that $\bar{E} \ll M$.

The statistical mechanical derivation of the density matrix (\ref{thermal}) makes use of the fact that the number $N_\nom$ of microscopic black hole states after the emitting radiation equals 
$$
{\rm dim}{\cal H}_{M-E_\nom} = N_\nom  = N e^{-\beta E_\nom},
$$
while the total number of states in ${\cal H}_\AA\otimes {\cal H}_\BB$ with total mass $M$ is given by 
\be
{\rm dim} \left({\cal H}_\AA\otimes {\cal H}_\BB\right)_{M} =\sum_\nom {\rm dim} {\cal H}_{M-E_\nom}= NZ.
\ee
Equation (\ref{thermal}) thus follows from the fact that each black hole state is occupied with the same uniform probability.

Let us now return to the firewall argument \cite{amps}. The initial condition that the black hole starts at $t=0$ in the maximally entangled state (\ref{maxmix}) heavily restricts the possible form of the density matrix at the later time $t=\tau$ of the black hole and the radiation that was emitted during that time. It was argued in \cite{amps} that, after letting the state evolve and emit an 
extra segment of radiation, the maximally mixed form is essentially preserved. 
Suppose that this is true. One then arrives at the following form of $\rho_\nm$:
\bea
\label{rhoamps}
{\rho}^{\mbox{\scriptsize \sc AMPS}}_\nm = 
\frac 1 {NZ} \,  \delta_\nm \, \mathbb{1}_{{}_{\! M\! \spc -\! \spc E_\nom}} .
\ee
Now comes the firewall argument: a density matrix of the factorized form (\ref{rhoamps}) is 
clearly inconsistent with the assumption that Alice can fall through the horizon without disturbance.
A direct way to see the problem is to compute the mutual information
\be
I_\abAB = S_\aaA + S_\bBB  - S_\abAB
\ee
 between the interior Hilbert space ${\cal H}_\aaA$ and the outside Hilbert space
${\cal H}_\bBB$. Using eqns (\ref{rhoamps}) and (\ref{thermal}), one finds
\bea 
\label{ampsentropy}
S^{\mbox{\scriptsize \sc AMPS}}_\abAB\!  
= \log(NZ), \quad \qquad 
S_\AA   
= \, \log N - \betaH \bar{E}, \qquad \quad
S_\BB  = 
\log Z + \betaH \bar{E},
\eea
where $\bar{E} = \sum_n \pp_n E_n$ is the average amount of energy contained in the late radiation. Notice the appearance
of the extra term $\log Z = - \beta F$, with $F$ the free energy of the radiation.\footnote{For a free photon gas, $\beta F =- \beta \bar{E}/3$, leading to $S_\BB \! =\! (4/3)\beta \bar{E}$.  The fact that the entropy in the radiation is a factor $4/3$ bigger than the change $\Delta S_\BH = -\beta \bar{E}$ in the BH  entropy was  
first noted in \cite{zurek}.} We will comment on its significance in a moment.

Equation (\ref{ampsentropy}) tells us that
the black hole interior and the late radiation share zero mutual information 
$I^{\mbox{\scriptsize \sc AMPS}}_\abAB = S_\aaA + S_\bBB  - S^{\mbox{\scriptsize \sc AMPS}}_\abAB = 0.$
A local vacuum state near the black hole horizon, on the other hand, encodes measurable entanglement between modes on each side.
The density matrix (\ref{rhoamps}) evidently does not. 
The authors of \cite{amps} argue that this discrepancy is a logical consequence of  quantum information theory: 
 since all black hole degrees of freedom are entangled with the early radiation, and entanglement can only be shared once,
there is no room left for entanglement between the black hole interior and the late radiation. This contradiction led the authors of
\cite{amps} to conclude that the horizon of an old black hole can not be a smooth region of space time.

The above version of the firewall argument is not completely precise. In particular, the maximally mixed state (\ref{rhoamps}) is
not consistent with unitarity.  To see this, note that the  entropy $S^{\mbox{\scriptsize \sc AMPS}}_\abAB\! = \log NZ$ of the 
state (\ref{rhoamps}) is larger, by an extra term of $\log Z$, than the BH entropy $S_\BH 
= \log N$ of the initial state (\ref{maxmix}). The origin of the extra term is easily understood. The time evolution takes place in 
Hilbert space ${\cal H}^{\rm late} = {\cal H}_A \otimes {\cal H}_B$ which  is much bigger than the interior Hilbert space ${\cal H}_A = {\cal H}_\BH$ by itself.
The initial state (\ref{maxmix}) thus spreads out into an enlarged Hilbert space, and its coarse grained entropy gains an extra contribution proportional to the 
associated Helmholtz free energy $F$.
Unitary time evolution, on the other hand, preserves microscopic entropy. So (\ref{rhoamps}) can not be exactly equal to  
the density matrix  (\ref{maxmix}) evolved to a later time $t= \tau$. There must be off diagonal corrections that restore the unitarity.

We thus have found a small gap in the reasoning of \cite{amps}. In \cite{amps} the black hole and emitted Hawking radiation combined were treated as if
it forms a closed system, with a Hilbert space of some fixed size $N$. However, as the above simple analysis indicates, such a framework is too  restrictive to describe the Hawking emission process. Eqns (\ref{thermal}) and (\ref{rhonm}) only make sense as relations on a tensor product space ${\cal H}^{\rm late} = {\cal H}_{A} \otimes {\cal H}_{B}$ on which the $b$-modes can act independently.  The local environment of the black hole, into which it slowly releases radiation and information, has a separate Hilbert space that is spanned by states that do not evolve from the black hole states themselves. A radiating black hole is not a closed system, but an open system, just like a radiating atom.
Monogamy of entanglement holds in closed systems, but becomes less restrictive in open systems.

\subsection{Evaporation in the Interaction Picture}

Clearly, we need a more precise framework to model the evaporation of the black hole states (\ref{young}) and (\ref{maxmix}) in a way that 
manifestly preserves unitarity. We have already characterized a Hilbert space. We also need to characterize our Hamiltonian that describes the time evolution.
We will choose to describe the time evolution in terms of the Schwarzschild time $t$, the time coordinate used by an outside observer.  

We  first consider time evolution on the full Hilbert space, including the early radiation. 
In accordance with the decomposition (\ref{hilbtotal}), we write the Hamiltonian $\hH$ as 
\be
\hH= \hH_\BH + \hH_\R+ \hH_{int}, 
\ee
where $\hH_\BH$ acts only on interior black hole states, $\hH_R$ governs the evolution of the radiation outside the stretched horizon. $\hH_{int}$ represents the interaction Hamiltonian, that is responsible for the Hawking radiation process. 
The precise form of  all terms in the Hamiltonian depend on unknown  micro-physical details.
We will therefore make only minimal assumptions about the specific form of each term. All that we need to know is that $H_\BH$ and $H_\R$ each act within the respective Hilbert spaces, and that $H_{int}$ acts on the tensor product in a way that
allows the two sectors to reach a quasi-static thermal equilibrium.

We can now use the familiar interaction picture, in which the time dependence of operators is governed by the free Hamiltonian
 $\hH_\BH+ \hH_\R$, and states evolve through the interaction Hamiltonian $\hH_{int}$.  Following standard time-dependent perturbation
 theory, we may then expand the time-dependent states $\bigl|\Psi(t)\ra$  as linear combinations, with  time dependent coefficients, of {\it static} eigenstates of the non-interacting Hamiltonian. The static states are of the form $\bigl| \, j \, \ra |\, \nom \, \ra$ where
 $\bigl|\, j \, \ra  \in {\cal H}_\BH$ is a microstate of a non-radiating eternal black hole of mass $M-E_{\nomt}$, 
 that exist in the fiducial reference world in which the interaction Hamiltonian has been turned off\footnote{In other words, $\bigl|\, j \, \ra$ is assumed to be an eigenstate of $H_\BH$ with eigen value $M-E_\nom$.}, and $\bigl|\, \nom \, \ra \in {\cal H}_R$ represents a static radiation state with energy $E_\nom$ that lives
 outside the stretched horizon of the static eternal black hole. This sitation is depicted in the left diagram in fig. 1.
 
\begin{figure}[t]
\begin{center}
\includegraphics[scale=.3]{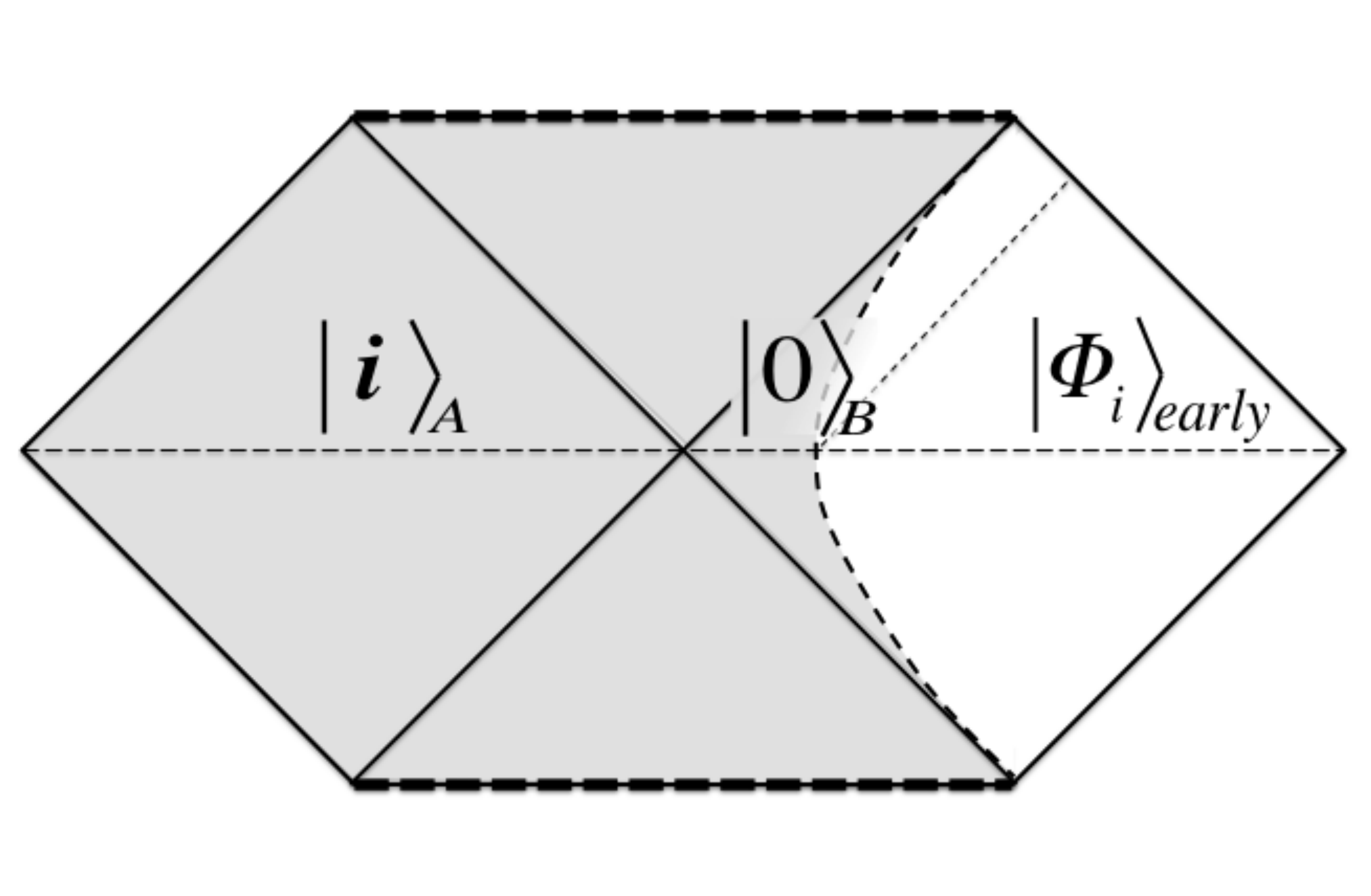}
\caption{\small
{A stationary element of the Hilbert space on which the interaction Hamiltonian acts is a tensor product of an eternal black hole state 
$\bigl|\,i\,\bigr\rangle_{A}$ with a state representing the radiation outside the stretched horizon. The radiation state at $t=0$ factorizes into a product of a state $\bigl|\Phi_i \bigr\rangle_{\rm early}$ 
containing the early radiation sand the vacuum state $\bigl|\,0\,\bigr\rangle_B$ of the late radiation.}}
\end{center}
\vspace{-0.5cm}
\end{figure} 

 Following \cite{amps}, we assume that the time evolution of states in ${\cal H}_R$  are well described by ordinary quantum field theory on curved space-time, and that all interactions with the
black hole microstates take place at the stretched horizon. In the context of the interaction picture this means that the time evolution of the radiation field can be represented in the static black hole background. Thus, all complications of the evaporation process, including the backreaction, are contained in the time evolution 
of the quantum states generated by the interaction Hamiltonian $\hH_{int}(t)$.

To proceed, we need to take slightly better look at the radiation states $\bigl|\Phi_i\bigr\rangle$. Superficially,  $\bigl|\Phi_i\ra$ 
specifies initial conditions only for radiation on a spatial slice outside the stretched horizon. However, not all modes in the outside region need to pass through this slice. As indicated in fig.~1, the stretched horizon is a semi-permeable membrane from which right-moving modes can emanate.
Hence, the Cauchy data need to include initial conditions at or inside the future stretched horizon.  The natural and only consistent choice is to impose vacuum boundary conditions, as measured in the Schwarzschild time, for the outgoing modes at $t=0$. 
In this way we guarantee that the mass $M-E_\nom$ of the eternal black hole is all contained in the interior part $\bigl|\, j \, \ra$ of the wave function.

From a semi-classical space-time perspective, it looks perhaps a bit suspect to assume that the radiation modes start out in their ground state as measured in the static frame, 
since this represents a singular state to an infalling observer. It is important to keep in mind, however, that smoothness of the
horizon is not an input assumption of our set-up. Our basic principle is to preserve the rules of quantum mechanics, and to describe the evaporation process in the same
way as one would describe the emission of a photon by an atom. With this starting point, Hawking radiation is not a consequence of a geometric regularity condition, but the result of turning on the interaction Hamiltonian $H_{int}$. From this perspective, the fact that the late radiation starts in the vacuum state means that one is dealing with spontaneous emission, rather than stimulated
emission. 

With this motivation, we postulate that the quantum states of the radiation at $t=0$  factorize into the tensor product of state $\bigl|\Phi_\iii \bigr\rangle_{\rm early}$, describing the
early radiation, times a vacuum state $\bigll|\,0\,\bigrr\rangle_\bB$ that specifies the initial condition for the late radiation.
\be
\bigll|\spc   \Phi_\iii \bigrr \rangle_{\! \eE}\,=\bigll|\,0\,\bigrr\rangle_\bB\;
\bigll|\spc 
 \Phi_\iii \bigrr \rangle_{\rm early}. \ee
Here $\bigl|\, 0\,\bigrr \rangle_b$ denotes the Boulware vacuum of the late radiation.  
The state $\bigll|\Phi_i\bigrr\rangle_{\rm early}$  is stationary, since the early radiation is assumed to have already left the interaction region at $t=0$. In other words, the early state is entirely decoupled from the late evolution. In the following we mostly forget about $\bigl|\Phi_i\bigr\rangle_{\rm early}$. All we need to know is that after tracing $\rho = |\Psi\ra \la \Psi|$ over  ${\cal H}_R^{\rm \, early}$ we end up with a maximally mixed state of the form (\ref{maxmix}), multiplied with the vacuum density matrix  $\bigl|0\ra_\bB \la 0|$ for the late radiation.

We are interested in the time evolution of the quantum state on ${\cal H}^{\rm late} = {\cal H}_\AA \otimes {\cal H}_B$, describing the interior black hole together with the late radiation, starting from $t=0$. It is sufficient to consider this time evolution  for pure initial states 
$\bigl|\,i\, \bigr\rangle |\,0\,\bigr\rangle$. This state evolves  through the interaction hamiltonian $H_{int}$, and thus after time $\tau$  it
takes the form  
$$
\sS(\tau) \spc \,\bigll|\,\iii\,\bigrr\rangle \bigll|\,0\, \bigrr\rangle_\bB
$$
where $\uU(\tau)$ denotes the evolution operator
\be
\uU({\tau})= T\! \exp\,\Bigl(- \frac{i}{\hbar}\!\int_0^{\tau} \!\!\!\! dt\, \hH_{int}(t)\Bigr).
\ee
Since the interaction Hamiltonian couples the black hole interior to the radiation space ${\cal H}_\bB$, the time evolved state
decomposes as
\be
\label{timestep}
\sS(\tau) \spc
\,\bigll|\,\iii\,\bigrr\rangle \bigll|\,0\, \bigrr\rangle_\bB \,=\, \sum_{\nomt, \,\jjj }\, 
C_{{ \nomt,\,\jjj}}^{\,\; \iii} \;   \bigll |\,\jjj \,\bigrr\rangle\,  
\bigll | \nom \bigrr \rangle_\bbbb .
\ee
The coefficients $C_{{ \nomt,\,\jjj}}^{\,\; \iii}$ represent the time-dependent probability amplitude for finding the black hole in the lower energy state $\bigll|\,j\,\bigrr\rangle$, combined with Hawking radiation in the state $\bigll| \,\nom\,\bigrr \rangle{}_{\bbB}$, at the time $t=\tau$.  
The states $\bigll|\,j\,\bigrr\rangle$ in (\ref{timestep}) are black holes with energy $M-E_\nom$, which span a Hilbert space
of dimension ${\rm dim}\spc {\cal H}_{M-E_\nom} = e^{-\beta E_\nom} N$. So for given $\nom$, the transition amplitudes $C_{\nom,j}^{\; i}$ 
combine into a non-invertible matrix. 

In the following, it will be convenient to write the $C_n$ coefficients as explicit matrix elements  $C_{\nomt, j}^{\, i} = \bigll\langle\spc j\spc\bigrr|\, \cC{}_{\nomt}\, \bigll|\spc i \spc \bigrr\rangle$ of operators 
acting on the black hole Hilbert space.\footnote{Note that, since  $\uU(\tau)$ acts on the tensor product ${\cal H}_\AA \otimes {\cal H}_\BB$, 
the rhs is still operator valued on ${\cal H}_A$.}
\bea
\label{Cn_def}
\cC{}_{\nomt}\,\equiv \, \raisebox{-3pt}{${}_\bbbb$}\bigll\langle \spc \nom\, \bigrr|  \uU(\tau) \, \bigll| \,0\,\bigrr \rangle_\bbbb
\ee  
Equation (\ref{timestep}) can then be succinctly written as (here we simplify $\uU(\tau)$ to $\uU$)
\bea
\label{ntimestep}
\sS\bigll|\,\iii\,\bigrr\rangle\bigll| \,0\,\bigrr \rangle = \sum_{\nomt }\, 
\cC_{{ \nomt}}\;   \bigll |\,\iii \,\bigrr\rangle \, 
\bigll | \spc \nom \, \bigrr \rangle . \ \
\eea
The unitarity constraint $\uU^\dag \uU = 1$ implies that
\bea
\label{unitarity}
\sum_{\nomt,\, \jjj} C_{{ \nomt, \,\jjj}}^{\,\, \iii} \spc\,  C_{{ \, \nomt, \,\jjj}}^{\ast\, k}\ =\ \delta^{\iii k}, \quad & &  \quad 
\sum_\nomt \cC^{\,\dag}_\nomt \cC_\nomt \, =\, \mathbb{1}_{{M}} .
\eea
Here and in what follows, $\cC_n$ and $\cC_n^{\, \dag}$ will always denote maps from ${\cal H}_{{M}}$\!  to  ${\cal H}_{{M-E_n}}$ and back.
For the rest, quite little is known about the $C_{{ \nomt, \,\jjj}}^{\,\, \iii}$ coefficients: their precise form depends on secret details of the micro-physics. 
Our main assumption will be that they look like large ergodic matrices \cite{fastscrambler}, with statistical properties consistent with  black hole thermodynamics.
We will see that this will enable us to determine their coarse grained structure with sufficient accuracy to decode, with the help of some familiar quantum information technology 
\cite{preskilllectures}, the local semi-classical physics on both sides of the black hole horizon.
Equation (\ref{ntimestep}) will form the starting point for much of the rest of our discussion.

Our main story is quite independent of the detailed form of the interaction Hamiltonian $\hH_{int}(t)$. But for concreteness and to gain some intuition, it may be useful to have a specific model in mind. Suppose that the radiation is described by a free field $\phi$. One of our postulates states that the black hole and radiation interact only at the stretched horizon. Locality then implies that $\hH_{int}(t)$ is expressed in the radiation field $\phi(t)$, and its first derivative, at the stretched horizon. This leads to a simplified model of the form 
\be
\hH_{int}(t) = \int \!\! dt \, \bigl(\phi_{\rm in} (t)\spc \vV_{\rm in}(t) + \phi_{out}(t) \vV_{out}(t)\bigrr). 
\ee
The interaction contains both incoming and outgoing fields, and hence describes black hole formation and evaporation.  Here we focus on the evaporation process, which just involves the outgoing field $\phi_{out} = \phi$. 
The corresponding interaction operator $\vV(t)$ can be expanded in modes $\vV_\omega$ and $\vV^{\,\dagger}_\omega$, that map black hole states with mass $M$ to states with mass $M\pm\omega$. The matrix elements
$
V_{ij}(\omega)= \bigl\langle\, j\,\bigrr|  \vV_\omega\bigll|\, i \ra\, 
$ determine the emission and absorption rates of particles of frequency $\omega$. Semi-classically this process is described by the creation of a Hawking pair. Hence it is natural to conjecture that the operators $\vV(\omega)$ and $\vV^{\,\dagger}(\omega)$ have something to do with the creating or annihilating particles behind the horizon.

In principle one can express the $\cC_{\nom}$ operators in terms of the$\vV$ operators as follows: one inserts the Fock states of the radiation and commutes the annihilation operators through the evolution operator. One obtains an expression involving products of a number of $\vV$ operators, one for every emitted particle. It is suggestive that the number of $\vV$ insertions matches the number of  infalling partner modes of the Hawking pairs.  This suggests that it is natural to view the transition amplitudes $\cC_{\nom}$ as creating a state with a collection of particles behind the horizon.  Our aim is to make this intuition more
precise. 

There exists of course a close parallel between the above discussion and the GKPW dictionary that underlies the AdS/CFT correspondence \cite{maldacena}. 
Indeed, we may compare
the process of falling into a black hole with entering  the near-horizon geometry of a stack of D-branes. In this comparison, the AdS space replaces the black hole interior
geometry, the CFT represents the interior Hilbert space, the value of the scalar field 
$\phi$ at the stretched horizon plays the role of the asymptotic boundary value of the bulk field in AdS, and the $\vV$ operators represent the CFT operator dual to $\phi$.

\subsection{Black Hole Entanglement Revisited}

As a first application of our set up, let us revisit the time evolution of the young and old black hole states (\ref{young}) and (\ref{maxmix}).

First consider the young black hole. At time $\tau$ it has evolved into a state of the form
\be
\label{newyoung}
\rho^{\spc \rm young}_\abAB = \sS\bigl|\, i\, \ra|\, 0\, \ra_\bB \la \, 0 \, \bigr| \la \, i \,\bigrr|\sS^{\, \dag},  
\ee
which we may expand in terms of the radiation basis as in equation (\ref{rhonm}), with
\bea
\label{nnyoung}
\rho^{\spc \rm young}_\nm \is  \cC_\nom \spc \bigll |\, i\, \ra \la \,i \,\bigrr|\cC^{\, \dag}_\mom. 
\eea
We now make the assumption that the  amplitudes $\cC_n$ are large ergodic matrices. In practice this means that we can view their matrix elements as being randomly chosen from an appropriate ensemble. The properties of this ensemble are completely fixed by quantum statistics, and should be consistent with the usual laws of thermodynamics.

In particular, if we start with a generic pure black hole state, the density matrix on the radiation states is expected to take the thermal form (\ref{thermal}). Thus statistically one should have the following identity
\be
\label{stats}
\la \,i \,\bigrr| \cC^{\, \dag}_\mom  \cC_\nom \bigll |\, i\, \ra =
\pp_n\, \delta_{mn} ,
\ee
with $\pp_n$ the normalized Boltzmann weight~(\ref{thermal}). 
This should be viewed as a coarse grained equality that holds for generic states $\bigl|\, i\, \ra$.  

Let us check this relation by writing the left hand side explicitly in components 
\be
\label{statts}
\,\sum_{j} C_{\nom,j}^\iii C^{* \, \iii}_{\mom,j}= \pp_n\, \delta_{mn}, 
\ee
and use our knowledge about the size of the black hole Hilbert spaces. First take $\nom = \mom$: we are then summing $Ne^{-\beta E_\nom}$ positive terms -- so the answer is clearly non-zero -- and since all amplitudes are statistically of equal size, the result must be proportional to $e^{-\beta E_\nom}$. The unitarity condition (\ref{unitarity}) tells us that the sum over $\nom$ of the entire expression is equal to $1$, so this fixes the normalization.  Next suppose that $\nom \neq \mom$. Then we are summing of the order of $N e^{-\beta E_n}$ terms with random phases. This sum is strongly suppressed relative to the diagonal term. The relative suppression of the off-diagonal terms is of order $1/\sqrt{N}$.
We thus find that statistically the above identity (\ref{statts}) indeed holds.

Before proceeding to the old black hole, let us determine the mutual information between the radiation and the black hole state. Since the initial state is pure, the entropy the black hole interior $A$, the radiation region $B$, and their union $AB$, is easily calculated, and gives
\bea
\label{youngentro}
{\rm young \ BH}:\ \quad  \ \ \ \ \  \begin{array}{cc}  S_\abAB = 0 , \quad &  \quad \  S_\BB = \log Z + \beta \bar{E},  \qquad\qquad \\[3mm]
S_\AA =  S_\BB, \quad & \ \ \ I_\abAB = 2S_\BB. \qquad\qquad \end{array} 
\eea
We see that the radiation region is maximally entangled with the internal black hole and carries the thermal entropy, as expected for a near horizon state. Indeed, on information theoretic grounds,
there seems to be no obstruction to identify the young black hole state (\ref{newyoung}) with a smooth local vacuum state. We will make this identification explicit in the next section.

Now let us look at the old black hole (\ref{maxmix}). After time $\tau$ it has evolved into the state 
\bea
\label{newrhoold}
 \rho_\abAB^{\spc \rm old}\spc \is \, \, \uU\, \rho(0)\spc \uU^{\, \dag}, \quad\mbox{with}   \qquad \rho(0)= \frac{1}{N}\mathbb{1}_M \otimes |\spc 0\spc \ra_\bB \la \spc 0 \spc\bigr| 
\eea
The density matrix is obtained from that of the young black hole (\ref{newyoung})by averaging over all initial black hole states. It can be expanded in the radiation basis as in (\ref{rhonm}) with
\bea
\label{rhonmold}
 \rho^{\spc \rm old}_{\nom\mom} \is % \frac{1}{N} 
 \, \frac 1 {N} \,
 \cC_\nom \cC^{\, \dag}_\mom . 
\eea
The fact that the density matrix $\rho_B$ of the radiation region $B$ is again thermal follows from $
  \frac 1 {N}\, \tr_\AA \bigl(  \cC^{\, \dag}_\nom \cC_\mom\bigr) = \pp_n\, \delta_{mn}$
which is automatically obtained from (\ref{stats}) by averaging over all black hole microstates.

Let us again compute the entropies associated with the three regions $A$, $B$ and $A \cup B$. First we note that the density
matrix (\ref{newrhoold}) manifestly satisfies  
\be
\bigr(\rho_\BH^{\spc \rm old} \bigr)^2 = \frac{1}{N} \rho_\BH^{\spc \rm old},
\ee 
from which we immediately see
that it still carries entropy $S_\abAB = \log N$, as it should. 
This result should be contrasted with the entropy $S_\abAB^{\mbox{\scriptsize \sc AMPS}}  = \log NZ$ 
of the diagonal mixed state (\ref{rhoamps}). The relation between the two density matrices can be understood if we introduce a phase averaging
procedure, that averages over all possible phases of the transition amplitudes $C_{j,\nom}^{\, \iii}$. Performing this phase average eliminates all
off diagonal terms of $\rho^{old}_{\abAB}$, resulting in the maximally mixed density matrix $\rho^{\mbox{\scriptsize \sc AMPS}}$
\bea
\label{rhoav}
\overline{\rho^{\, \rm old}_{\mom\nom}} \; =\, \frac 1 {{N}}\,   \overline{ \cC_n  
\cC^{\, \dag}_m} \, \is \, 
\frac 1 {NZ} \,  \delta_{nm}\, \mathbb{1}_{{}_{\! M\! \spc -\! \spc E_n}} .
\eea
The extra term $\log Z$ in the entropy is thus due to coarse graining as a result of taking the phase average. One should keep in mind that there are many off-diagonal 
corrections to  (\ref{rhoav}). Individually, these are all small but there are many of them.
So collectively, they can have a substantive effect.

The diagonal form (\ref{rhoav}) is still appropriate for computing the entropy of the black hole interior region $A$. We thus find the following result for the entropies
and mutual information
\bea
\label{oldentro}
{\rm old\  BH}: \ \ \qquad  \ \ \ \  \begin{array}{cc}  
S_\AA   
= \, \log N - \betaH \bar{E}, \quad & \quad S_\abAB = \log N \qquad\qquad \\[3mm]
S_\BB  = \spc
\log Z + \betaH \bar{E}  , \quad  & \quad  I_\abAB{} =    \log Z. \qquad \qquad \end{array} 
\eea
We see that the radiation region $B$ and the interior region $A$ now do share mutual information!
This means that there is a non-zero amount of entanglement between the two.

How is it possible that the maximally mixed state (\ref{maxmix}) evolves into a state with non-zero mutual information between region $A$ 
and $B$? Clearly, this would not be possible if we would think of Hawking particles as quantum bits that were previously contained inside
the black hole. The key point is that the state (\ref{maxmix}) is embedded inside a larger Hilbert space that, besides the black hole states, 
also contains the radiation modes.  Since monogamy of entanglement does not hold in open systems, the time evolution can
generate new entanglement.  Note however that, comparing with eqn (\ref{youngentro}), the amount of shared
information is a bit less than for the young black hole.
For a free photon gas,
$\log Z = \beta E/3$, so the amount of entanglement for the old black hole is roughly a factor of 8 less than for the young black hole.
The question we will need to investigate is if this amount of shared entropy is sufficient to describe an approximate vacuum state near the horizon.

Above we have used quantum statistical arguments to determine sums of products of transition amplitudes $C_{n,j}^i$ and their complex conjugates.
We will use this same reasoning repeatedly in the following sections. One may try to formalize this using a random matrix approach. One practical choice would
be to pick a gaussian  matrix ensemble, where one can use Wick's theorem. A more accurate characterization of the 
matrix ensemble is as the space of all  matrices $C_{n,j}^i$ that satisfy the unitarity constraint (\ref{unitarity}).
The language of Wick contractions still serves a useful terminology for labeling  and estimatingthe different resonant contributions.
 We will not try develop this formalism further here, since for the cases of interest the relevant result can be determined by simple reasoning as above.

\section{Passing through the Horizon}

\vspace{-2mm}

In this section we will take a more detailed look at this question. Our goal is to first  make explicit 
how a relatively young quantum black hole, given either by a pure initial state (\ref{young}) or a partially entangled quantum state {\it  can} in fact describe a smooth horizon state.
From a quantum information perspective, this result is not a real surprise.. Nonetheless, it is still an important first step. It has up to 
now been far from obvious how {\it any} quantum mechanical model of a  black hole can be consistent with having a smooth horizon. 
This logical tension is at the heart of the black hole information paradox.

Specifically,  we will show how the transition amplitudes $\cC_\nom$ can be used to reconstruct a mirror of the QFT
Hilbert space, spanned by the Hawking pairs of the outside radiation modes. Our construction makes use of the close
analogy between our description of the evaporation process and the loss of coherence of a quantum computer \cite{haydenpreskill,preskilllectures}.
In the case of a quantum computer, one can recover the lost quantum information via the use of
quantum error correction (QEC).  
This new tool will then enable us to investigate the issues raised by the information paradox, and the firewall argument in more quantitative terms.

\subsection{Quantum Error Correction}

Many of the quantities and relations  introduced above have direct cousins in the 
quantum information context.  Let us briefly mention a few entries of the dictionary \cite{quantuminformation,preskilllectures}.

One can think of the black hole as a quantum storage device exposed to environmental noise. 
Eqn (\ref{ntimestep}) then describes the time evolution, the leakage of quantum information from the device into the environment,
written in the so-called Kraus representation. The transition matrices $\cC_\nom$ are known as Kraus operators.
The emission of a Hawking quantum corresponds to the occurrence of an error, and  the Kraus operators are typically denoted~by~$\textit{\textbf E}_\nom$.
We will continue to use the notation $\cC_\nom$. The evolution operator $\sS = \sum_\nom \cC_\nom \otimes  |\nom\ra_\bB \la 0|$ is known as the error super operator,
and the image of the $\cC_\nom$ operators is called an error subspace.

Not all states in the Hilbert space can be simultaneously protected from error.
A QEC code includes the specification of  a code subspace, the space of encoded states that one wishes to protect.  We will
denote states in the code subspace by
\bea
\bigl|\spc \ibar \spc \ra \in {\cal H}_{\rm code}.
\eea
The code subspace is chosen independently of the actual state of the system.  In the black hole context, one may be tempted to try to
identify the whole interior Hilbert space with the code subspace, but this is not realistic. The presence of extra states and degrees of freedom outside
the code space is essential, so that one can build in sufficient redundancy to safeguard the encoded information.  

We choose the encoded states
to span a subspace within the Hilbert space ${\cal H}_M$ of mass $M$, of dimension ${N_{\rm code}}\ll N$. We will comment later on the corrections
that come into play when we try to take the limit ${N_{\rm code}} \to N$.

Not all types of quantum errors can be corrected. A necessary and sufficient condition is that for any pair of Kraus operators $\cC_\nom$ and $\cC_\mom$, 
and for any pair of states $\bigl|\spc \ibar \spc \ra$ and $\bigl|\spc \kbar \spc \ra$ in the code subspace, one has \cite{quantuminformation,preskilllectures} 
\bea
\label{keyrel}
\la\smpc \ibar \smpc \bigl| \spc {\cC^{\, \dag}_\nomt \cC_\momt} \spc \bigl|\spc \kbar \spc \ra \is \pp_\nom \, \delta_{\nom \mom}\, \delta_{\ibar \! \kbar}.
\eea
Here we fixed the normalization via eqn (\ref{stats}). If the property (\ref{keyrel}) does not hold, then errors can ruin the distinguishability of 
different encoded states and quantum information could be irrevocably damaged. For us,  (\ref{keyrel}) is more than just a pre-condition.
As we will see, it is the key algebraic identity
that will enable us to decode the interior Hilbert state and exhibit the semi-classical horizon geometry.

In the black hole context, the condition (\ref{keyrel}) is naturally
satisfied: it generalizes eqn (\ref{stats}) to the case where $\bigl|\spc i \spc \ra$ and $\bigl|\spc k \spc \ra$ are different states, and follows from
the same considerations outlined below eqn (\ref{stats}).  It expresses the fact that $\cC_\nom$ and $\cC_\mom$ are statistically
 independent complex random matrices, that only interfere constructively when all terms in the component expansion of the product are of
 the form $\bigl|C_{n,j}^i|^2$. Equation (\ref{keyrel}) should therefore not be read as an exact identity, but as an equation that for generic states holds with
 very high accuracy.  Its corrections are suppressed by a relative factor of $1/\sqrt{N}$. 
However, even while essentially true for any pair of states in ${\cal H}_\AA$, eqn (\ref{keyrel}) can not be read as a true global statement of the form ${\cC^{\, \dag}_\nomt \cC_\momt} = \pp_\nom\spc \delta_{\nom\mom}$, since this would appear to imply that  
${\cC^{\, \dag}_\nomt \cC_\nomt}$ is an invertible matrix, which it is not:  its image has dimension
$N_\nom = N e^{-\beta E_\nom} < N$. So for given $E_\nom$, we can use eqn (\ref{keyrel}) with confidence as long as we choose
$$
{N_{\rm code}} \ll N_\nom=Ne^{-\beta E_\nom}.
$$ 
We will encounter the necessity of this inequality at many other places.

A quantum error correcting code is specified by the pair $\bigl({\cal H}_{\rm code}, \{ \cC_\nom\} \bigr)$, the code subspace and the collection of errors it tries to correct.
Given eqn (\ref{keyrel}), one can naturally associate a notion of entropy to a given QECC, given by $S_{\rm code} = \sum_\nom \pp_\nom \log \pp_n$

\def\overlinde{{}}

Let us now describe the error correcting operation. It makes use of the recovery operators 
\be
\label{recover}
R_\nom = \frac 1 {\sqrt{\pp_\nomt}} \sum_{\ibar} \bigl|\ibar \ra \la \ibar \bigr|  \cC^{\; \dag}_\nom.
\ee
These operators $R_\nom$ can be combined into a single super-operator $\rR$
with the help of an `ancillary' Hilbert space ${\cal H}_\aaaa$,  or simply  `the ancilla', spanned by basis states $\bigl|\spc\nom\spc\ra_\aaaa$ in one-to-one
correspondence with the basis states $\bigl|\spc\nom\spc\ra_\bbbb$ of the late radiation.
Hence ${\cal H}_\aaaa$ is isomorphic to ${\cal H}_\BB$. 
As an example, in case the radiation field is described by a free scalar field $\phi$, the ancilla ${\cal H}_\aaaa$
is the Fock space spanned by creation and annihilation
operators ${\bf a}^\dag$ and ${\bf a}$.

The recovery super-operator $\rR$  acts on the tensor product ${\cal H}_A \otimes {\cal H}_\aaaa$ via
\be
\label{superoperator}
\rR \, \bigr| \, j \, \ra \bigl| \spc 0\spc \ra_{\! \aaaa} \spc = \sum_\nomt R_\nom \, \bigl|\, j \, \ra \bigr| \, \nom \, \ra{}_{\nspc \aaaa}
\ee
 The ancillary Hilbert space is an important ingredient of quantum error correction, since as we will see shortly, 
 it provides the depository into which it dumps the entanglement  between the code space and the environment \cite{preskilllectures}. 
We emphasize, however, that in our context -- unlike for the usual application of QEC codes --
 the ancillary Hilbert space  ${\cal H}_\aaaa$ is just an intermediate algebraic device, that
 helps us put the recovery operator in a convenient form.  Since our main purpose is to keep track of information flow,
it is therefore important keep in mind that  the ancillary Hilbert space never really exists and
 needs to be projected out in the end, in a way that manifestly preserves unitarity.
 We will return to this point, once the physical meaning of the ancilla has become clear.

The recovery operator is designed to reverse the error: applying $\rR$ to a time evolved state $\uU \bigl|\, \iii \, \ra \bigl|0\ra_\aaaa$, one recovers the original state
$\bigl|\, \iii \, \ra$ times some state on the Hilbert space ${\cal H}_\BB \otimes {\cal H}_\aaaa$, provided that the initial state $\bigl|\, \iii\, \ra$ lies inside the
code subspace.
The verification of this claim is straightforward, and makes use of the property (\ref{keyrel}) of the $\cC_\nom$'s:\footnote{Here we keep the short-hand notation $\uU$ and $\rR$ for the extensions $\uU \otimes \mathbb{1}_\aaaa$ and $\rR \otimes \mathbb{1}_\bbbb$ of the 
superoperators to the tensor product ${\cal H}_A \otimes {\cal H}_\aaaa \otimes {\cal H}_\BB$.}
\bea
\label{recoveryc}
\rR \spc \sS \, \bigl| \, i \, \ra \bigl| \spc 0 \spc \ra_\aaaa \bigl| \spc 0\spc \ra_\bbbb   \! & = &  \sum_{\mom,\nom} R_\mom \cC_\nom \bigl|\, i \, \ra \bigl| \, \mom \, \ra_{\! \aaaa} | \, \nom \, \ra_{\nspc \bbbb} \nonumber \\[2.5mm]
\is \sum_{\mom,\nom,\kbar}  \frac 1 {\sqrt{\pp_\momt}}\, \bigl|\kbar \ra \la \kbar \bigl|  \cC^{\, \dag}_\mom  \cC_\nom \bigl|\, i \, \ra \bigl| \, \mom \, \ra_{\nspc \aaaa} \bigl| \, \nom \, \ra_{\nspc \bbbb}  \\[3mm]
 \is \sum_{\kbar} |\kbar \ra \la \kbar  \bigl|\, i \, \ra\;  \sum_{\nom} \sqrt{\pp_\nom}\, \bigl| \, \nom \, \ra_{\! \aaaa} \bigl| \, \nom \, \ra_{\nspc \bbbb} .\nonumber \eea
We read off that
\bea
\label{recoveryr}
\rR\spc  \sS\,  \bigl| \, i \, \ra \bigl| \spc 0 \spc \ra_{\! \aaaa}  \bigl| \spc 0\spc \ra_{\nspc \bbbb}  & = &  \left\{\begin{array}{cc} \bigl|\, i \, \ra  \,  \bigl|\, 0_{\spc U}  \ra \qquad & {\rm if}\ \, \bigl| \, i \, \ra \in {\cal H}_{\rm code}  \\[4mm] \ 0 \qquad \ & {\rm if}\ \,\bigl| \, i \, \ra \notin {\cal H}_{\rm code} \end{array}\right.
\eea
where $\bigl|\, 0_{\spc U}\ra$ is the following familiar looking state, defined on the tensor product of the Hilbert space of the 
outside radiation region with the ancillary Hilbert space
\be
\label{rindlers}
\bigl|\, 0_{\spc U}\ra = \sum_\nom \sqrt{\pp_\nom} \; \bigl|\spc \nom \spc \ra_{\nspc \aaaa} \spc \bigl|\spc \nom \, \ra_{\nspc \bbbb}.
\ee
Given that $\pp_n$ are the Boltzmann weights, we recognize this state as the Unruh vacuum state, with the ancilla states playing the role of the interior radiation region.
We see explicitly how the $\rR$ operation works: it sweeps all entanglement between the time evolved black hole and the emitted radiation, 
 encoded in $\uU\bigl| \spc i \spc \ra   \bigl| \spc 0\spc \ra_{\nspc \bbbb} $, into entanglement of the radiation with the ancilla.
 Thus the recovery operation restores the purity of the black hole state, provided it started out within the code space.

We see from (\ref{recoveryr}) that the recovery operator $\rR$ is not a unitary map, since it projects out all states that do not map back into the code space. 
This means that the operation $\rR$ as defined in (\ref{superoperator})-(\ref{recover}) in fact is not really a valid superoperator.
Preferrably, one would want the recovery operation to be a unitary operation. To achieve this, consider the operator
\be
\label{pidef}
\pPi  \spc =\spc  \raisebox{-.5pt}{${}_\aaaa$}\nspc \la \spc 0 \spc \bigl| \smpc \rR^{\spc \dag} \rR\spc  \bigl| \spc 0\spc \ra_{\nspc \aaaa} .
\ee
We may also write $\pPi = \sum_\nom R^\dag_\nom R_\nom =  \sum_{\ibar,\nom} \frac 1 {\pp_\nom} \,\cC_\nom\spc  \bigl|{} \ibar{} \ra \la{} \ibar{} \bigl|  \cC^{\, \dag}_\nom$.
If $\rR$ had been a valid superoperator, then $\Pi$ would have been equal to the identity operator. Instead it defines a projection operator on the
space of all states that can be reached by acting with the $\cC_\nom$ operators on the code subspace. In other words, it is the projection of
all states that the code states can evolve into as a result of time evolution over time $\tau$. 

A straightforward calculation, similar to the one done
in equation (\ref{recoveryc}), shows that $\pPi$ indeed has the property of a projection operator:
\be
\pPi^2 = \pPi.
\ee
To complete the recovery super operator, one thus adds one extra term to the right-hand side of (\ref{superoperator}) proportional to the projection operator
$\mathbb{1} - \pPi$ on the orthogonal complement.
\bea
\bigl( \mathbb{1} - \Pi \bigr)  \spc \uU \bigl|\, i  \, \ra  \bigl| \spc 0\spc \ra_{\nspc \bbbb}  & = &  \left\{\begin{array}{cc} 0   \qquad & {\rm if}\ \, \bigl| \, i \, \ra \in {\cal H}_{\rm code}  \\[4mm] \ \uU \bigl|\, i  \, \ra  | \spc 0\spc \ra_{\nspc \bbbb} \qquad \ & {\rm if}\ \,\bigl| \, i \, \ra \notin {\cal H}_{\rm code} \end{array}\right.
\eea
We see that the QECC has indeed succeeded in its task of protecting all states within the code subspace from loss of coherence. 

\subsection{Construction of the Interior Operators}

We have seen that the combined operation $\rR\spc \uU$, time evolution followed by quantum error correction, associates to every black hole initial state $\bigl| \, i \,\ra \in {\cal H}_{\rm code}$ an Unruh state of the form (\ref{rindlers}). Since for any state $\bigl| \, i \,\ra $, there is some  recovery operator $\rR$ for which $\bigl| \, i \,\ra $ lies within its code subspace, we have shown that $\bigl| \, i \,\ra $
(when evolved forward in time)  always contains a factor that can be decoded into a local
vacuum state near the horizon.  

We now define interior operators, that form the Hawking partner of the outgoing radiation modes. Specifically,
we associate to every operator on ${\cal H}_\BB$ a partner operator that acts  
on the interior black hole states. Our construction works for every state in the code subspace. 
Suppose that the black hole together with the radiation at time $t=0$ are described by some mixed state of the form
 \be
 \label{rhosmall}
 \rho(0)\; = \; \sum_{\ibar,  \kbar} \ \rho_{\ibar\!\kbar} \, \bigl| \spc \ibar \spc \ra \la \spc \kbar \spc \bigl|\   \otimes\ \bigl|\spc 0\spc \ra_\bB \la \spc 0 \spc\bigr|,
 \ee
 where the sum runs over some small subspace ${\cal H}_{\rm small}$ of the total black hole Hilbert space. We are then free to choose a code subspace 
such that ${\cal H}_{\rm small} \subset {\cal H}_{\rm code}$.  At time $t = \tau$, the state (\ref{rhosmall}) has evolved into the density matrix 
 \be
 \label{rhotau}
 \rho(\tau) = \uU \rho(0)  \spc \uU^{\, \dag}.
 \ee
 We wish to show that this density matrix describes a black hole with a smooth horizon.
 
 \def\bfA{ \textit{\textbf A}}
 
We will now define the interior operators. In the previous section, we introduced the ancillary Hilbert space ${\cal H}_\aaaa$, as a carbon copy of the
outside radiation Hilbert space ${\cal H}_\BB$. 
For us, this ancilla does not really exist, but is just introduced as a helpful algebraic tool. Even so, we are
free to interpret ${\cal H}_\aaaa$ as representing all states on the hidden Rindler wedge, the spatial region on the other side of the 
horizon from ${\cal H}_\BB$. Now
consider some operator $A$ that acts on ${\cal H}_\aaaa$. We wish to identify $A$ with the Hawking partner of an identical operator
acting on the outside radiation space ${\cal H}_\BB$. But we can not use the ancilla as our Hilbert space, so we need to transform $A$ into an operator
acting on the actual internal Hilbert space of the black hole.
We now associate to $A$ the following operator acting in the interior black hole
Hilbert space 
\bea
\label{adef}
 \textit{\textbf A} = \spc {}_\aaaa\nspc\la \spc 0 \spc \bigl| \smpc \rR^{\spc \dag}\nspc A\spc \rR\spc  \bigl| \spc 0\spc \ra_{\! \aaaa}.
 \eea
where $\rR$ is the recovery superoperator defined in (\ref{recover})-(\ref{superoperator}).
 Note that this is a vacuum expectation value in ${\cal H}_\aaaa$. So  $\textit{\textbf A}$ is a proper linear operator acting on ${\cal H}_\AA$, 
 the state space of the black hole, and the ancilla is indeed just a virtual device.
Eqn (\ref{adef}) is modeled after the definition (\ref{pidef}) of the projection operator onto the 
 code subspace. Indeed,  $\Pi = \spc \raisebox{-.5pt}{${}_\aaaa$}\nspc  \la \spc 0 \spc \bigl| \smpc \rR^{\spc \dag}\nspc \mathbb{1}_a \spc \rR\spc  \bigl| \spc 0\spc \ra_{\! \aaaa}$ assumes the role of the identity operator on the interior Hilbert space ${\cal H}_\aaaa$.

As an example, in case the
radiation field is described by a free scalar field $\phi$, we can choose to consider $A = \phi(y)$, where $y$ is a point on the left Rindler wedge
behind the horizon, that is, $\phi(y)$ is defined as a mode expansion in terms of ancillary oscillators $a^\dag$ and $a$.
The general definition (\ref{adef}) then associates an effective local field operator $\mbox{\boldmath $\phi(y)$}$, acting on the
interior black hole  Hilbert space, via 
 \be
 \label{phidef}
 \mbox{\boldmath $\phi(y)$} =  \spc {}_\aaaa\nspc\la \spc 0 \spc \bigl| \smpc \rR^{\spc \dag}\nspc \phi(y) \spc \rR\spc  \bigl| \spc 0\spc \ra_{\nspc \aaaa}.
 \ee
 Eqns (\ref{adef})-(\ref{phidef}) are our proposal for a microscopic realization of the interior QFT operators. 
 
 To support this identification, 
we first need to verify that the map (\ref{adef}) between $A$ and $\bfA$ is an isomorphism between operator product algebras. In other words,
for any two operators $A_1$ and $A_2$, we need to show that
\be
\label{opcheck}
\bfA_1 \spc \bfA_2 = \spc \raisebox{-.2pt}{${}_\aaaa$}\!\smpc \la \spc 0 \spc \bigl| \smpc \rR^{\spc \dag}\nspc A_1 \spc A_2 \spc \rR\spc  \bigl| \spc 0\spc \ra_{\!\smpc \aaaa}.
\ee
This equation tells us that applying the map (\ref{adef}) after taking the product gives the same result as taking the product after applying the map.  
If this holds, then the $\bfA$ operators indeed have the same operator product algebra as the effective QFT operators.

Let us check that (\ref{opcheck}) indeed holds. \footnote{In this section we assume that ${N_{\rm code}}\ll N$. We will look at the case ${N_{\rm code}} \to N$ in the next section.}
It is useful to expand out the definition (\ref{superoperator}) of the recovery superoperator. The definition (\ref{adef}) can be expanded as
\be
\label{newdef}
 \textit{\textbf A} \is  \sum_{\nom, \mom}\,   R^\dag_\nom \,A_\nm \, R_\mom ,
 \ee
 where $A_\nm = \spc  {}_\aaaa\!\smpc\la \spc \nom \spc  \bigl| \spc A\spc \bigl| \spc \mom\spc  \ra_{\!\smpc \aaaa}$.
 We see that the combination $R^\dag_\nom R_\mom$  takes over the role of the operator $\bigl|\spc \nom \spc \ra_\aaaa \la \mom |$ in the ancilliary Hilbert space. 
As a free field example: with this notation, we may define the {\it effective} interior creation and annihilation
operators ${\bf a}^\dag$ and ${\bf a}$ via 
$$
{\bf a}^\dag = \sum_{\nom,\mom} R^\dagger_\nom \la \nom \bigl|\spc a^\dag \bigl|\mom\ra R_{\mom}\quad \mbox{and} \qquad {\bf a} = \sum_{\nom,\mom} R^\dagger_\nom  \la \nom \bigl|\spc a \spc \bigl|\mom\ra R_{\mom}.
$$ 
The ${\bf a}$ and ${{\bf a}^\dag}$ operators act purely within the internal black hole Hilbert space ${\cal H}_\AA$, and thus commute with the creation and annihilation modes $b^\dagger$ and $b$ of the radiation. 

With the definition (\ref{newdef}), the requirement (\ref{opcheck}) that $\bfA$ satisfies the QFT operator product algebra amounts to the identity 
\bea
\label{pialgebra}
R^\dag_\nom R_\som\, R^\dag_\tom R_\mom \spc \is \, \delta_{\som\tom} \, R^\dag_\nom R_\mom.
\eea
As shown in the Appendix, this property indeed holds with great accuracy, as long as ${N_{\rm code}} \ll N$.
The precision with which it is valid is a measure for the successfulness of the quantum error correcting procedure $\rR$. For us, it
measures the faithfulness with which we can recover the operator algebra of the interior low energy effective field theory. It is a strength of our
formalism that we can in principle compute the corrections and limits of validity of the effective theory;
we will consider the correction to the operator algebra when discussing the limit ${N_{\rm code}}/N \to 1$ in the next section.

\begin{figure}[t]
\begin{center}
\includegraphics[scale=.48]{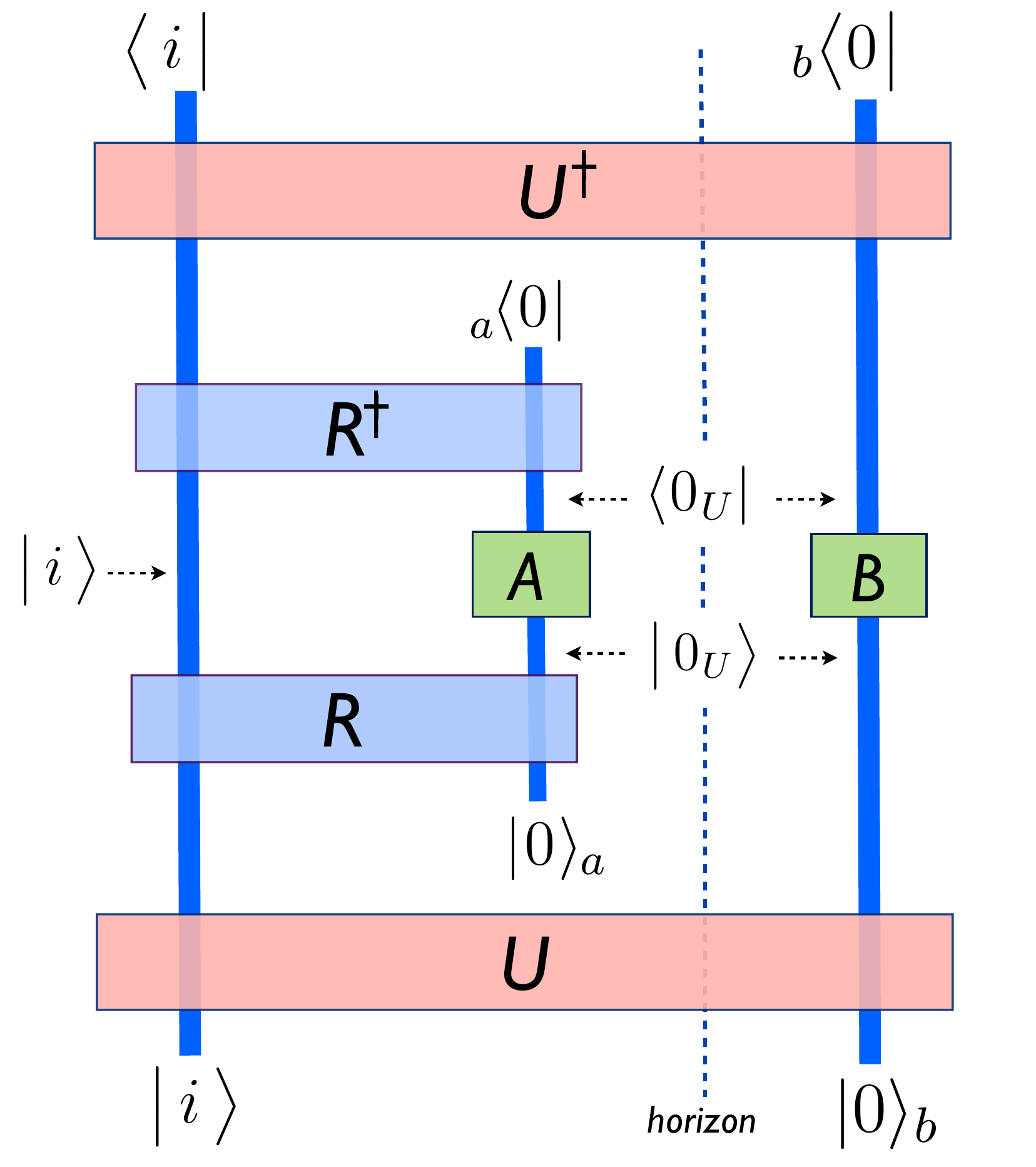}
\caption{\small
{Steps in the quantum computation of the expectation value $\tr(\rho \bfA B)$ for a pure state density matrix $\rho(0) = \bigl|\spc i\spc \ra\la\spc i\spc \bigr|$. 
The initial state $\bigl|i\ra \bigl|0\ra_b$ first evolves via the evolution operator 
$\uU$. The interior operator then acts via a recovery operation $\rR$  followed by the operator $A$ acting on the ancilla. The outside operator
$B$ acts directly on the time evolved radiation state. At the instant that $A$ and $B$ act, the interior state is identical to the initial state $\bigl|i\ra$, and
the $a$ and $b$ states are in the local vacuum state $\bigl|0_U\ra$.
Then follows the conjugate of the recovery operation ${}_\aaaa\la 0 \bigl| \rR^\dag
$, which eliminates the ancilla. Finally, one projects back onto the final state ${}_b\la0\bigl|\la i \bigl| \uU^\dag$.
}}
\end{center}
\vspace{-0.5cm}
\end{figure}

Next we compute the expectation value of the operators in the state $\rho(\tau)$ given in (\ref{rhotau}). Here we assume that the
initial state (\ref{rhosmall}) at $t=0$ is contained inside the code subspace, and that ${N_{\rm code}}\ll N$.
 Let $A$ and $B$ denote some general operators in terms of the $a$ and $b$-modes, respectively, and let $\bfA$ be the interior operator
 defined via (\ref{adef}). The computation of the expectation value $\tr(\rho(\tau) \bfA B)$ is straightforward, and summarized in the Appendix. 
 We find that the expectation value coincides with that of the effective field theory in the local Unruh vacuum state
\bea
\label{unruhvev}
\tr\bigr(\rho(\tau) \spc  \textit{\textbf A} B \bigr) \, =\,  
\la \spc 0_{\spc U} \bigr| A\spc B \bigl|\, 0_{\spc U}\ra .
\eea  
This is the main result of our paper. It shows that the black hole state can look like a smooth horizon state, without straining the rules of quantum information theory.

Another way to obtain this result is to use the expansion (\ref{newdef}) in terms of matrix elements. In the Appendix it is shown that
\be
\label{rhobb}
 \tr_\aaA\! \bigl(\rho \, R^\dag_\nom R_\mom \bigr) \, =\,  \sqrt{\pp_\nom \pp_\mom }\, \, \bigl|\spc \nom \, \ra_\bbbb \la\spc \mom \, \bigr|  .
\ee 
From this one directly computes  (here $B_{\nom\mom} =  {}_\bbbb\!\smpc\la \spc \nom \spc  \bigl| \spc B\spc \bigl| \spc \mom\spc  \ra_{\!\smpc \bbbb}$)
\bea
\label{rindelrv}
\tr \bigll(\rho \, \bfA \, B \bigrr) \is \sum_{\nom,\mom} A_\nm \tr\bigll(\rho \, R^\dag_\nom R_\mom  \, B \bigrr) \, = \, \sqrt{\pp_\nom \pp_\mom}\,  A_\nm B_\nm.
\eea
This is the expectation value in the local Unruh vacuum (\ref{rindlers}).

What made this magic work? The definition (\ref{adef}) of the interior modes involves a QECC operation, which requires
detailed knowledge of the emission amplitudes $\cC_n$, whose precise form depends on unknown details of 
 Planck scale physics. It looks like cheating to assume that Alice can access these observables,
 without actually having such detailed knowledge. The situation is of course similar to other
 holographic dualities, like AdS/CFT.  From the dual perspective, the interior space-time emerges in a mysterious way. But for Alice, 
 the smooth local space-time is a classical reality, relative to which the operators (\ref{adef}) behave like ordinary observables.
  So she has no trouble using them.  Note, however, that Alice  does  not organize the modes in terms of their Schwarschild energy $\omega$,
 but  reassembles them into local Minkowski modes, that make no reference to the precise location of the horizon.

To make this a bit more concrete, consider the two point function of local field operators, say one behind and one in front of the horizon.
The general result (\ref{unruhvev}) gives that
\bea
\tr\bigr(\rho \,  \mbox{\boldmath $\phi_a$}(y) \phi_\bbbb(x) \bigr) \, =\,  
\la \spc 0_{\spc U} \bigl| \phi_\aaaa(y) \spc \phi_\bbbb(x) \bigl|\, 0_{\spc U}\ra ,
\eea  
where the subscript indicates that $\phi_a$ is made up from $a$-modes, and  $\phi_\bbbb$ from $b$-modes.
What is the meaning of this expectation value?  Strictly speaking, it does not yet describe what we want. In our discussion so far, we have only 
included the outgoing modes that are emitted by the black hole. The ancillary $a$ modes play their guest role as the Hawking partner
of the outgoing $b$ modes: they initially live on the left region in fig 1, but can be analytically continued to live just behind the future horizon.
To really get the complete set of modes behind the horizon, we would need to extend our framework a bit, by adding the infalling modes.
Since the horizon state looks smooth, the infalling modes  can be analytically continued to the region behind the horizon. So we have
indeed explained how a QFT mode, and by extension any object, can fall into a black hole.

Note that we can also try to go beyond the past horizon, by switching  time direction.
 The outgoing modes behind the past horizon are obtained by analytic continuation of outgoing modes on the right region in fig 1.
 The time-flipped Hawking partners of the incoming modes live behind the past horizon, and are obtain by analytic continuation
  of modes on the left region in fig 1. To define them, we would need to introduce transition amplitudes
  for the ingoing modes involved in the black hole formation process.

\subsection{Ancilla versus Black Hole Final State}
\vspace{-2mm}

Our use of the ancillary Hilbert space ${\cal H}_\aaaa$ is somewhat reminiscent of the black hole final state proposal of \cite{finalstate}. However, there is a key difference:
unlike the final state proposal, our procedure is completely in accord with the conventional rules of quantum mechanics. We have stated explicitly in what Hilbert space
we are working, and time evolution is unitary.  Our internal operators (\ref{adef})] involve an expectation value inside an auxilary Hilbert space ${\cal H}_\aaaa$, but we could have avoided that altogether by directy using the definition (\ref{newdef}). So
we work at all times in the same Hilbert space, without tensoring in or projecting out any extra degrees of freedom.

The black hole final state proposal of \cite{finalstate} works differently. There the ancilla ${\cal H}_\aaaa$ and the radiation space ${\cal H}_\bbbb$  
are new Hilbert space sectors that arise out of nowhere whenever a back hole gets formed, and are postulated to start out in the state 
$\bigl|0_U\ra = \sum_\nom \sqrt{\pp_\nom} \bigl|\nom\ra_\aaaa \bigl|\nom\ra_\bbbb$. 
\be
\label{hmcrime}
\bigl| \, i \, \ra    \rightarrow \bigl| \, i \, \ra  \bigl|0_U\ra.
\ee
This is a crude prescription, however, that immediately obscures the rules of quantum mechanics. Where was 
the Hilbert space ${\cal H}_\aaaa \otimes {\cal H}_\bbbb$ before the black hole was formed?
How did it end up in a maximally entangled state $\bigl|0_U\ra$ with the radiation modes? How can one keep track of quantum information if such things can happen?

Superficially, our equation (\ref{recoveryr}) looks very similar to (\ref{hmcrime}). For us, however, equation (\ref{recoveryr}) does not represent time evolution,
and the state  $\bigl| \, i \, \ra  |0_U\ra$ does not exist in our physical Hilbert space ${\cal H}_\AA \otimes {\cal H}_\bbbb$. 
So let us interpret the proposal of \cite{finalstate}
in the same spirit. In equation (\ref{hmcrime}), we should not think of the ancilla states $\bigl|\nom\ra_\aaaa$ as real, but virtual.  Indeed we  can undo the damage done by 
the rash introduction of the extra Hilbert space and restore unitarity by writing\footnote{This equation follows from $\uU \bigl| \spc i \spc \ra \bigl|\spc 0 \spc \ra_b = \Pi U  \bigl| \spc i \spc \ra \bigl|\spc 0 \spc \ra_b = 
{}_a\la 0 \bigr| \rR^\dag \rR \uU  \bigr| \spc i \spc \ra \bigl|\spc 0 \spc \ra_a \bigl|\spc 0 \spc \ra_b = {}_a\la 0 \bigl| \rR^\dag \bigr| \, i \, \ra  \bigl| 0_U \ra $.}
\be
\label{finproj}
\uU \bigl| \spc i \spc \ra \bigl|\spc 0 \spc \ra_\bbbb  = {}_\aaaa \la 0\bigl| \rR^\dag \bigl| \, i \, \ra  \bigl| 0_U \ra .
\ee
The left-hand side is a manifestly unitary evolution of the initial state $\bigl| \spc i \spc \ra \bigl|\spc 0 \spc \ra_\bbbb$, while the right-hand side looks like a non-unitary final
state projection ${}_\aaaa \la 0 \bigr| \rR^\dag$ applied to the state $\bigl| \, i \,  \ra  \bigl|0_U\ra$. The apparent non-unitarity of the final state prescription cancels out
against the initial unitarity violation of introduction of the additional Hilbert space sector. 

This interpretation immediately circumvents the main substantive objection against the final state proposal made in \cite{preskillgottesman}, where it was pointed out that 
any non-trivial dynamics between the two sectors would immediately destroy the coherence of the physically observable state. But as
the above discussion makes clear, the ancillary Hilbert space should be viewed as dynamically completely decoupled from the physical Hilbert space:
the evolution operator $\uU$ and Hamiltonian $H$ acts only on ${\cal H}_\AA \otimes {\cal H}_\bbbb$ and leaves ${\cal H}_\aaaa$ inert. 
That the dynamics exactly factorizes in this way would look very unreasonable if ${\cal H}_\aaaa$ had been physically real, but is obvious in our set-up.
The final state proposal is only salvageable --   but then loses much of its  explanatory potential -- by realizing that the ancillary Hilbert space 
${\cal H}_\aaaa$ is an artificial construct.

If the ancilla states are not real, how then are we are able to still employ quantum error correcting technology?
QEC involves the reversal of decoherence. When the code block interacts with its environment, the two become entangled and 
the Von Neumann entropy of both systems increases, as seen from (\ref{youngentro}) The recovery operation $\rR$ implements an
entanglement swap: it purifies the code space by sweeping all its entanglement with the radiation modes into the ancilla.
Hence reality of the ancilla is absolutely essential for quantum error correction, since where else can the entanglement entropy go? 

The resolution of this apparent conflict is as follows. As seen from the definition (\ref{adef}), $\textit{\textbf A} \! =\!  {}_\aaaa\nspc\la \spc 0 \spc | \smpc \rR^{\spc \dag}\nspc A\spc \rR\spc  | \spc 0\spc \ra_{\! \aaaa}$,
the interior operators make only temporary use of  the recovery procedure. Reading from right to left, they tensor in the auxiliary ground state $\bigl| \spc 0\spc \ra_{\! \aaaa}$, 
and perform the recovery operation $\rR$.  This step removes entropy from the code block and temporarily stores it inside ${\cal H}_a$. Then it applies $A$,  and instantly undoes the recovery 
by acting with the `final state projection' ${}_\aaaa\nspc\la \spc 0 \spc \bigr| \smpc \rR^{\spc \dag}$. This final state projection -- which is necessary for restoring unitarity --
erases the ancilla and dumps the entropy back into the interior Hilbert space ${\cal H}_\aaaa$.\footnote{This entropy transport goes hand in hand with transport of heat.
In light of Jacobson's arguments \cite{jacobson}, relating entropy and heat flow with the GR equations, it would clearly be very interesting to study this in more detail.}
So the QEC procedure is able to perform an entanglement swap, but at a price: the
 interior black hole Hilbert space space clogs up with entropy as time passes by. 

This looks like a recipe for trouble. As the entropy of the black hole density matrix increases, one needs a bigger and bigger code space for 
performing the recovery procedure on the whole density matrix. Eventually ${N_{\rm code}}$ starts approaching the size $N$ of the full Hilbert 
space. But as we have seen, the quantum error correcting code becomes less and less 
reliable with increasing ${N_{\rm code}}/N$, and as a result, our identification of the interior effective QFT operators starts breaking down. 
This is a manifestation of the firewall problem.

\subsection{Fidelity and Errors}
\vspace{-2mm}

  Why do quantum error correcting codes have any useful application in the study of black hole information and entanglement?
In the context of their application to quantum computers, QEC codes are designed to protect the quantum coherence of a particular part of the Hilbert space, the code subspace, 
by correcting the errors that occur by emission of particular quanta. It does so by undoing the transformation that is responsible for the emitted quanta of radiation,  and moving the entanglement with these quanta to another part of the Hilbert space that is outside the code subspace. 

In the black hole context, the quantum  recovery operation acts in an exactly identical way on the Hilbert states, but the QEC code plays a rather different role \cite{haydenpreskill}. 
The code subspace  is  typically already in some highly entangled with  the early radiation outside the black hole. The error correcting procedure still works by undoing the transformation due to the emission of radiation. If it succeeds, it entangles the radiation again with another part of the Hilbert space. As we
have just shown, this entanglement helps preserve the entanglement across  the smooth the horizon.

In any QEC procedure there is always a small but finite chance that it fails. In the situation of  quantum computer in an approximate pure state,
the error will result into some partial decoherence: a small part of the code subspace will  become entangled with the outside. As a result, the corresponding 
internal quantum state  is no longer suitable for doing quantum computations. In the black hole application, failure of the QEC code means the opposite. Occurrence of an error
means that the code did not preserve the entanglement between the code subspace with the early radiation. Therefore the QEC error will cause a swap in entanglement:
it releases a qubit of information from the code subspace  and lets it escape to the outside. This is how the information leaks out of the black hole \cite{haydenpreskill}.

What is the fidelity and how big is the error of the QECC? To determine
this, we need to look at how well the recovered state approximates the original state.
A good way to compute this error is to determine how well the projection operator $\pPi$, when evolved backwards in time, approximates the
identity operator on the code subspace
\bea
\label{errorcheck}
\Pi  \spc \uU \bigl|\spc \ibar  \spc \ra  \bigl| \spc 0\spc \ra_{\nspc \bbbb} \is \uU \bigl|\spc \ibar  \spc \ra  \bigl| \spc 0\spc \ra_{\nspc \bbbb} \; + \ \mbox{error}. %\ldots
\eea
This equality is without error for a perfect QECC with $100\%$ fidelity. Plugging in the definitions for the left-hand side gives
\bea
%\Pi \spc \raisebox{0pt}{$\uU$} \spc  |\spc \ibar  \spc \ra  | \spc 0\spc \ra_{\nspc \bbbb}  = 
\sum_{\kbar, \nom,\mom}
 \frac 1 {\pp_\nom} \, \cC_\nom\spc  \bigl|\spc \kbar\spc \ra \la\spc \kbar\spc \bigl|  \cC^{\, \dag}_\nom \cC_\mom\smpc \bigl|\smpc \ibar \smpc \ra  \bigl| \spc \mom \spc \ra_{\nspc \bbbb}. 
\eea
We evaluate this expression via the same procedure as before: we keep only those terms in which every matrix element $C_{\nom, j}^{\, i}$ 
constructively interferes with its complex conjugate. In the limit ${N_{\rm code}}\ll N_\nom$,  the dominant interference
occurs between $\cC^{\, \dag}_\nom \cC_\mom$ for the term with $n=m$. This leads to the
first term on the right-hand side of (\ref{errorcheck}). This is the good term. 

\def\hH{\textit{\textbf H}} 
But there is also a bad term: when ${N_{\rm code}}/N_\nom $ becomes non-negligible, the other
interference term becomes important: $\cC_\nom^{\; \dag}$ on the right can constructively interfere with $\cC_n$ on the left. 
In this way we find that the error is given by
\be
\label{errorc}
\mbox{error}\; =\;  \Error \uU \bigl|\spc \ibar  \spc \ra  \bigl| \spc 0\spc \ra_{\nspc \bbbb},\quad\mbox{with}\qquad \Error\,=\,\sum_\nom\, \frac{{N_{\rm code}}}{N_\nom}\; \mathbb{1}_{M-E_\nom}.
\ee
%\bea
%\label{errorc}
%\ldots \; =  \;   {\mathbf\frac{{N_{\rm code}}}{Ne^{-\beta\textit{\textbf H}}} \, 
%\uU |\spc \ibar  \spc \ra  | \spc 0\spc \ra_{\nspc \bbbb}
%\eea
%with $1/Ne^{-\beta \hH } = \sum_\nom \mathbb{1}_{M-E_\nom}/ Ne^{-\beta E_\nom}.$ 
The operator $\Error$ will be called the `error operator'. 
Hence to have an accurate recovery procedure, we must choose the code subspace small enough so that this error is negligibly small.
Thus we again encounter the condition that ${N_{\rm code}} \ll N_\nom$.

\newpage

\subsection{Approaching the Firewall}

The firewall argument  concerns the limit in  which the black hole density matrix  approximates the maximally mixed state (\ref{maxmix}). 
There are several motivations to study this limit. As emphasized in \cite{vanraamsdonk}, 
two  space time regions can be connected if they are maximally entangled at a microscopic level. 
These arguments suggest that to be able to fall through 
the horizon without drama, one actually needs the maximal entanglement between the inside and the outside. 
 Other related perspectives are the interpretation 
of black holes as information mirrors \cite{haydenpreskill}and fast scramblers \cite{fastscrambler},   and Mathur's argument that, 
as far as quantum information goes, order one corrections are needed to the usual semi-classical picture.

Motivated by these ideas, we will now study what happens as one approaches
the maximally mixed limit.
We will still assume that the initial state $\rho(0)$ in (\ref{rhosmall}) fits inside a code subspace of size ${N_{\rm code}} < N$, which we parametrize via 
\bea
\label{rationc}
\frac{{N_{\rm code}}}{N} = e^{-\beta\mu}.
\eea
The maximally mixed case corresponds to setting $N_{\rm code}/N=1$, so
that $S_{\rm code} = \log N_{\rm code}$ saturates the Bekenstein-Hawking bound $S_\BH = \log N$. We will call this the firewall limit.

Based on our set up, we can try to give a somewhat more precise characterization of the density matrix of an old black hole, as follows.
As pointed out above, the quantum error correction procedure involved in reproducing the internal effective field  theory is not perfect,
but has a finite probability of producing an error.   
In the previous section, we saw that,  when considering operators that act on an internal state with energy $M-E_\nom$, this error is of the order
\be
\label{rationce}
\frac{N_{\rm code}}{N_n} = e^{\beta(E_\nom-\mu)}.
\ee
The occurrence of an error coincides with an event in which the black hole releases information about its interior state into the environment.
So an important first question is:  what is the expected maximal rate at which these errors can occur?  At the late stages of evaporation, 
the number of information carrying  quanta should exceed the number of quanta that increase the black hole entropy.  Therefore, the error rate 
eventually exceeds the frequency of good recoveries. Our proposed definition of the density matrix of an old black hole is that its entropy has grown 
to the level such that the probability that any given Hawking particle carries out a qubit of quantum information into the environment is of order 1.  
This amounts to an error rate  of about one qubit  per black hole crossing time $M$.  So in the following we will set our time step $\tau$ equal to one crossing time.

The errors of the QECC give rise to  corrections to the internal low energy effective field theory. Consider the operator combination $R^\dag_\nom R_\mom$ featuring in the definition (\ref{newdef}) of the internal  operators. The erroneous Wick contraction produces an extra contribution
\bea
\label{errorcont}
\overline{R^\dag_\nom R_\mom}  \is  \frac 1 {\sqrt{\pp_\nomt\pp_\momt}} \sum_{\ibar} \overline{\cC_\nom\spc  \bigl|\ibar\ra \la\ibar \bigl|  \cC^{\, \dag}_\mom} = \delta_\nm \spc e^{\beta(E_n - \mu)}  \mathbb{1}_{M-E_\nom}.
\eea
Here the bar indicates the phase average. The normalization of the right-hand side is determined by unitarity, c.f.  eqn (\ref{rhoav}).
Note that it indeed vanishes when $e^{-\beta\mu}\to 0$, i.e. when the code space becomes negligibly small.
This extra interference term deforms the algebra (\ref{pialgebra}), that underlies the isomorphism between the internal operator
algebras and that of the radiation field.  Thus we need to reinvestigate our definition of the interior operators, and pick the 
definition that best preserves the effective QFT description.

The most reasonable prescription is to introduce a normal ordered product $:\!\!R^\dag_\nom R_\mom\!\!: $, defined as the operator product with the 
phase averaged contribution $\overline{R^\dag_\nom R_\mom}$ subtracted. In other words, the normal ordered operators are traceless.
This recipe preserves the identity $
 \tr_\AA \bigl(\rho :\!\!R^\dag_\mom R_\nom \!\!:\bigr) \, =\,  \sqrt{\pp_\nom \pp_\mom }\, \, |\spc \nom \, \ra_\bbbb \la\spc \mom \, | ,$
which ensures that, if we keep our definition of the internal operators of the form (\ref{newdef}) but now as normal ordered expression $:\!\!\aA \!\nspc :$, 
 the expectation values of the effective QFT still look like those in the local vacuum state $\bigl|\spc 0_U\ra$.
So from the  semi-classical perspective, it looks like the horizon is still smooth -- in the sense that all expectation values are well behaved.

The error has not gone away, however: the normal ordering procedure still leads to a deformation of the operator product
algebra (\ref{pialgebra})  
\be
\label{deform}
:\!\nspc R^\dag_\nom R_{\som}\nspc \!:\;:\!\nspc R^\dag_\tom R_\mom\!: \, = \, \delta_{\som\tom} \spc \bigl(\, :\! \nspc R^\dag_\nom R_\mom \!:\, +  \, \delta_{\nom\mom} \spc e^{\beta ( E_\nom - \mu)}\spc \mathbb{1}_{M-E_n}\bigr).
\ee
To verify this relation, observe that the product of two normal ordered operators is not normal ordered; this leads to an extra contribution equal to $\overline{R^\dag_\nom R_\mom}$.
This extra term can be ignored as long as $\mu \gg E_\nom$. When the energy $E$ starts approaching the value of $\mu$, however, it starts to show up as a manifestation of the error of the QEC operation. Before we investigate the form of the deformation, let us determine how big or small it is.

What is the probability that the QEC procedure produces an error? Consider the operator $\Pi$ introduced in (\ref{pidef}). 
It can be expanded as $ \pPi = \sum_\nom :\!R^\dag_\nom R_{\nom}\! :$, and
in the low energy effective theory, it represents the unit operator $\mathbb{1}$. 
%In the microscopic theory, $\Pi$ is the projection onto the space spanned by all states that the code states can transition into over time $\tau$.
It  satisfies $\pPi^2 = \pPi$ provided that the recovery precedure works perfectly.  But once we include the possibility of errors, $\pPi$ is no longer
a perfect projection operator. Using (\ref{deform}), we find a correction term given by
\bea
\label{goodbad}
\pPi^2 = \pPi + \Error, \quad & & \quad 
\Error = 
 \sum_\nom e^{\beta (E_\nom - \mu)}\,  \mathbb{1}_{M-E_\nom} .
\eea
We assume that $e^{-\beta\mu}$ is adjusted so that  $\Error$ is still small.
We may then interpret  $\Error$ as the operator that produces the error in the recovery operation.

To determine the relative magnitude of the `good' first term and the `bad' error term, let us take their expectation value. For concreteness, we assume that
the black hole density matrix at time $t=0$ takes the form $\rho(0) = \frac{1}{{N_{\rm code}}} \sum_{\ibar} \bigl|\ibar\ra \la \ibar\bigr|$. We then have
\bea
\tr_\AA\bigl(\spc \rho(\tau) \spc \Pi\spc ) \! \is\! \sum_\nom \pp_\nom \, \, \bigl|\spc \nom \, \ra_\bbbb \la\spc \nom \, \bigr| = \frac{1}{Z}\, e^{-\beta H_b} 
 \\[3mm]
\tr_\AA\bigl(\spc \rho(\tau) \spc  \Error \,
) \is
\sum_\nom e^{-\beta\mu}\,   \bigl|\spc \nom \, \ra_\bbbb \la\spc \nom \, \bigr| 
= 
e^{-\beta\mu}\mathbb\, {1}_\BB.
\eea
The good term is the thermal density matrix $\rho_\BB$ of the radiation, and thus its trace is equal to 1. So it is normalized as a probability measure.
When there is a non-zero error rate, however, this normalization needs to be adjusted, to account for the fact that the QEC operation 
is no longer has  $100 \%$ fidelity. It seems reasonable to interpret the second equation as the relative probability that an error takes place
in the quantum channel. It tells us that
the errors have an infinite temperature: every state $\bigl|\nom\,\ra$ in the radiation Hilbert space has , regardless of its total energy $E_\nom$,  equal probability 
$p \simeq e^{-\beta \mu}$ to be the carrier of some irretrievable error. This probability increases linearly with the $N_{\rm code}$, as expected.

As we increase the size of the code space, we see that there is a natural cross over point, where the error probability becomes of order 1.
This crossover takes place  when the dimension of the code space reaches the critical value 
\bea
\frac{{N_{\rm code}}}{N}\is e^{-\beta\mu} \simeq \frac{1}{\dim{\cal H}_\BB}.
\eea
One may think of this as a detailed balance equation: it picks the saturation value for the size of the code subspace, such that the QEC procedure starts producing
approximately one error per crossing time. Notice that the right-hand side is a UV sensitive quantity: it depends on the size of the short-distance cut-off
of the effective field theory that describes the Hawking radiation\footnote{We implicitly assume that the Hilbert space ${\cal H}_\BB$ already comes with some
natural IR cut off, say, on the order of several times the Schwarzschild radius of the black hole.}  and on the location of the stretched horizon.

When the error correction works,  we get a semi-classical description of the evaporation process.
But when it fails it produces modifications to the low energy effective field theory. What do they look like?
Consider the identity (\ref{opcheck}) that verifies that the mapping (\ref{adef})-(\ref{newdef}) preserved operator product relations. 
We now find that the right-hand side receives an extra diagonal term, 
 %$\sum_\nom (A_1 A_2)_{\nspc \nom\nom} e^{\beta(E_\nom -  \mu)} \spc \mathbb{1}_{M-E_\nom}$.
  which we may think of as a result of normal ordering. We find that the operator product algebra is deformed according to
\bea
:\!\! \bfA_1\!\nspc :\, :\!\! \bfA_2\!\nspc : \is :\!\! \bfA_1 \bfA_2\!\nspc: \, +\,    \sum_\nom \spc (A_1 A_2)_{\nspc \nom\nom} \,  e^{\beta(E_\nom -  \mu)} \spc \mathbb{1}_{M-E_\nom}.
\eea
How would a low energy observer interpret this deformation?  It is tempting to look for a way to translate this algebra in pure effective field theory language, perhaps with some
non-locality.  However, the extra term is 
not of the form (\ref{adef}), and acts on a  bigger Hilbert space than ${\cal H}_{\rm code}$, the space we used for our re-construction of the effective QFT.
Because of this, we have not found a sensible way to answer this question.

Perhaps this is indeed the wrong question to ask.  The presence of the extra term is dictated by unitarity of
the underlying microscopic theory, and the microscopic mechanisms that are responsible for releasing the information from the black hole 
are likely to be not describable by semi-classical physics. In our QEC terminology, they produce error terms that in magnitude 
are of order one. Moreover, once the black hole entropy starts to decrease, the number 
of information carrying radiation modes must exceed the number of modes that increase the entanglement, the error rate  must
exceed the frequency of successful recovery operations. This indeed indicates that in this regime the semi-classical description can no longer be trusted.  
We are then led to conclude that when corrections to the semi-classical theory  becomes important,
the effective QFT disintegrates. 
This is a manifestation of the firewall problem, and perhaps the manifestation of the formation of an actual firewall or fuzzball. \cite{mathur-fuzzball-review}.

This last conclusion is still a bit premature, however. It is quite conceivable that the breakdown of the effective QFT is simply a consequence of
our attempt to capture a close-to-maximally entangled state of a large system in terms of a single semi-classical reality. Indeed, what we have
shown is that a black hole with small enough non Neumann entropy relative to its BH entropy bound,  i.e. with $e^{-\beta\mu}\ll1$, supports 
a semi-classical geometry with a smooth horizon. Since a maximally mixed black hole is an incoherent sum of black hole states with less than 
maximal entropy, we have shown that it represents an incoherent sum of different (in the sense that they are incompatible with each other) semi-classical states. 
Usually this means that the semi-classical states are the states that we measure.

\newpage

\section{Summary}

In this final section we give an executive summary of our main results. Consider the total quantum state of the black hole (with internal states $\bigl| \,j\, \ra \,$) after emission of quanta of late radiation (described by states $\bigl| \,\nom\, \ra$) 
\be
\label{beginstate}
\bigl|\,\Psi \,\bigr\rangle ={1\over\sqrt{N}}\sum_{\nom,i, j} C^{\,i}_{\nom, j} \,\bigl| \,j\,\bigr \rangle \,\bigl| \,\nom\,\bigr \rangle\, \bigl| \,\Phi_i\bigr \rangle. 
\ee
Here $\bigl|\Phi_i\ra$ is the state of the  early radiation. Here we assume that the late radiation started
coming out at some time $t=0$, and (\ref{beginstate}) is the state at some later time $t =\tau$. Since the early radiation is assumed to be dynamically
decoupled, we are allowed to treat the states $\bigl|\Phi_i\bigr\rangle$ as static.

 We would like to investigate with which part of the wave function the radiation state
$\bigl| \,\nom\, \bigr\rangle$ is entangled, the black hole state or the early radiation state. In \cite{amps} it is argued that, since for an old black hole 
the Hilbert space of the early radiation is so much larger than that of the black hole itself,  the radiation state must be maximally entangled with the environment state $\bigl|\Phi_i\bigr\rangle $ and has no entanglement with the black hole state $\bigl|\,j\,\bigr\rangle $.\footnote{ In section 2.4 we already established that this is not entirely true, even in the maximally entangled situation. The radiation state is always shares entanglement with the black hole. } This appears to exclude the existence of a smooth horizon. The question that we will first try to answer is:
\medskip

\noindent 
${}$~\parbox{16cm}{\it Under which conditions can an observer who has only access to the internal black hole states $\bigl|\, j\, \bigr\rangle$ accurately determine in which state $\bigl|\, \nom\, \bigr\rangle_b$ the late radiation has been emitted?}

\smallskip

\noindent
To focus the discussion, let us assume that  an outside observer has detected the radiation  in a particular state $\bigl|\, \nom\,\bigr\rangle$.  Thus instead of the full quantum 
state $\bigl|\, \Psi\,\bigr\rangle$ we now look at
\be
\label{newstatet}
\bigl|\!\nspc\spc\bigl|\nom \,\bigr\rangle\!\!\bigr\rangle = {1\over \sqrt{N_\nom}}\sum_{i, j} C^{\,i}_{\nom, j} \,\bigl| \,j\,\bigr \rangle \,\bigl| \,\nom\,\bigr \rangle\, \bigl| \,\Phi_i\bigr \rangle,
\ee
where we renormalized the state so that its norm is equal to one. 
Is there a measurement that acts on the states $\bigl|\,j\,\bigr\rangle $ whose outcome is going to tell us about the quantum numbers $\nom$ of the late radiation?

The key point on which our paper is based is that the coefficients $C^{\,i}_{\nom, j}$ are not just some arbitrary set of numbers: for given $\nom$ they are transition amplitudes
between black hole states, as a result of emitting the radiation in state $\bigl|\nom\ra$. Hence the $C^{\,i}_{\nom, j}$ are determined by the microphysics, and
represent the matrix elements of particular operators $\cC_\nom$, that {\it act on the interior part of the black hole Hilbert space}. We assume that the $C^{\,i}_{\nom, j}$ are all statistically independent random matrices.
Now pick some arbitrary state $\bigl|\spc \mom\spc \ra$, with the same energy as $\bigl|\spc \nom \spc \ra$, and consider
 the operator $\Pi_{\mom} = \frac{1}{\pp_\mom} :\!\! \cC_\mom \mathbb{1}_{\rm code} \cC^\dag_\mom\! :$ with matrix elements
\be
\label{pimats}
\left(\Pi_{\mom}\right)_{jk}= {1\over \pp_\mom} \sum_{\overline{i}} C^{*\, \overline{i}}_{\mom,j} C^{\, \overline{i}}_{\mom,k}  \, -\,
{\rm trace}
%  \frac{{N_{\rm code}}}{N_\mom} \delta_{jk}.
\ee
Here $\pp_\mom = \frac 1 Z e^{-\beta E_\mom}$ is the Boltzmann weight. 
%and $N_\mom = N e^{-\beta E_\mom}$ is the number of black hole state with mass $M-E_\nom$.
The sum over $\ibar$ runs over some code subspace of the black hole Hilbert space of dimension ${N_{\rm code}} < N$. We will now show that this operator, which only
acts on the black hole interior, is capable of recovering the information about the radiation state $\bigl|\, \nom\,\ra$. 

Applying this operator $\Pi_\mom$ to (\ref{newstatet}) gives 
\be
\label{piact}
\Pi_{\mom}\bigl|\!\bigl|\nom \,\bigr\rangle\!\!\bigr\rangle = {1\over \sqrt{N_\nom}}\sum_{i, j,k} C^{\,i}_{\nom, j} \left(\Pi_{\mom}\right)_{jk}  \,\bigl| \,k\,\bigr \rangle \,\bigl| \,\nom\,\bigr \rangle\, \bigl| \,\Phi_i\bigr \rangle.
\ee
Since the $C_{n,j}^i$'s are ergodic matrices, whose coefficients all arbitrary complex phases, we can evaluate this sum with the help of standard statistical reasoning. We need to look for the resonant contributions. Inserting (\ref{pimats}) into (\ref{piact}), we obtain an expression with a sum over two internal indices: $j$ and $\ibar$. The index $j$ 
labels black hole states of mass $M-E_\nom$ and thus runs over $N_\nom$ indices and $\ibar$ labels the code subspace and thus runs over ${N_{\rm code}}$ indices.
We evaluate both sums  by keeping only resonant terms, by using eqns  like (\ref{keyrel}) and (\ref{errorcont}).  
We call the sum over $j$ the `good contraction' and
the sum over $\ibar$ the `bad contraction'. The second term in (\ref{pimats}) is chosen such that it cancels the `bad contraction'.
The good contraction gives back the state $\bigl|\!\bigl|\nom \,\bigr\rangle\!\!\bigr\rangle $ when $\mom=\nom$, but with a projection on to the code subspace states 
\bea
\label{projectn}
\Pi_{\mom}\bigl|\!\bigl|\nom \,\bigr\rangle\!\!\bigr\rangle =\delta_{\mom\nom} \Pi_{\rm code}\, \bigl|\!\bigl|\nom \,\bigr\rangle\!\!\bigr\rangle, %+ {N_{code}\over N_\nom} \bigl|\!\bigl|\nom \,\bigr\rangle\!\!\bigr\rangle
\eea
%\qquad\mbox{where} \qquad
where $\Pi_{\rm code} = \sum_{\nom} \Pi_\nom$. So it looks like we have been
able to establish that the black hole state is maximally entangled with the radiation mode $\bigl|\, \nom\, \ra$, provided that the black hole part of the initial state lies in
the code subspace. So why can't we just take the code space equal to the complete interior Hilbert space and set ${N_{\rm code}} = N$?

There's the catch. Eqn (\ref{projectn}) is true as a coarse grained equation. In computing the sums in (\ref{piact}) we ignored many off-diagonal 
terms, which individually are all very small. We should worry, however, about their collective effect. One place where this enters is in the
verification that the operator $\Pi_\mom$ really acts like a projection operator with $\Pi_\mom^2 = \Pi_\mom$. By the same calculation method,
we find
\be
\label{newerror}
\Pi_\mom^2 = \Pi_\mom +\Error_\mom , \qquad \quad
\Error_\mom = \frac{{N_{\rm code}}}{N_\mom} \, \mathbb{1}_{M-E_\mom},
\ee
where $\mathbb{1}_{M-E_\mom}$ is the unit operator on the black hole Hilbert space with mass $M-E_\mom$. The extra term  $\Error_\mom$
the error of the recovery operation. So to have an accurate recovery without appreciable errors, we need that the code subspace is small.

We conclude  that the black hole is still maximally entangled with the radiation mode when the state of the black hole fits inside of some suitably chosen
code subspace with
\bea
\Pi_{\rm code}\bigl|\!\nspc\spc\bigl|\nom \,\bigr\rangle\!\!\bigr\rangle = %\!\!\bigr\rangle\, {}^{{}_{{}_{\,\textstyle\sim}}} \!\!\!\!\!\!\;\!{}_{{\textstyle\sim}}
\;\bigl|\!\nspc\spc\bigl|\nom \,\bigr\rangle\!\!\bigr\rangle\,\quad\mbox{and}\qquad {{N_{\rm code}}} \ll N_\nom.
\eea
In section 3.2 we show that in this case one can reconstruct the interior QFT observables in terms of the recovery operators. This is compatible
with information theoretic constraints: the reconstruction works only as long as the entanglement entropy with the early radiation is sufficiently 
smaller than the entropy $\log N_\nom$ of the space of the states $\bigl|\, j\, \bigr\rangle$.

\begin{figure}[t]
\begin{center}
\includegraphics[scale=.29]
{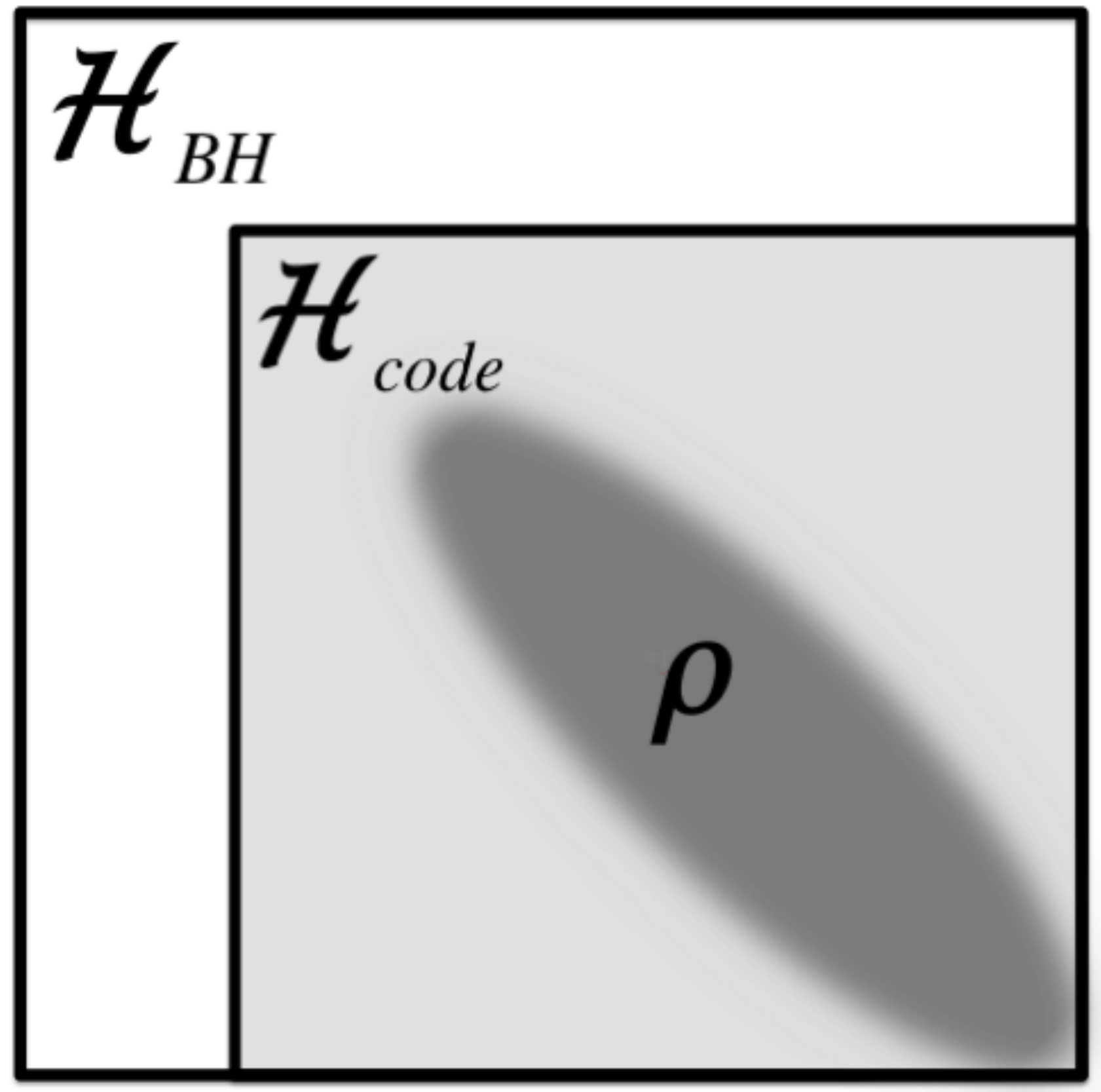}~~~~~~~~~~~~~~~\includegraphics[scale=.30]
{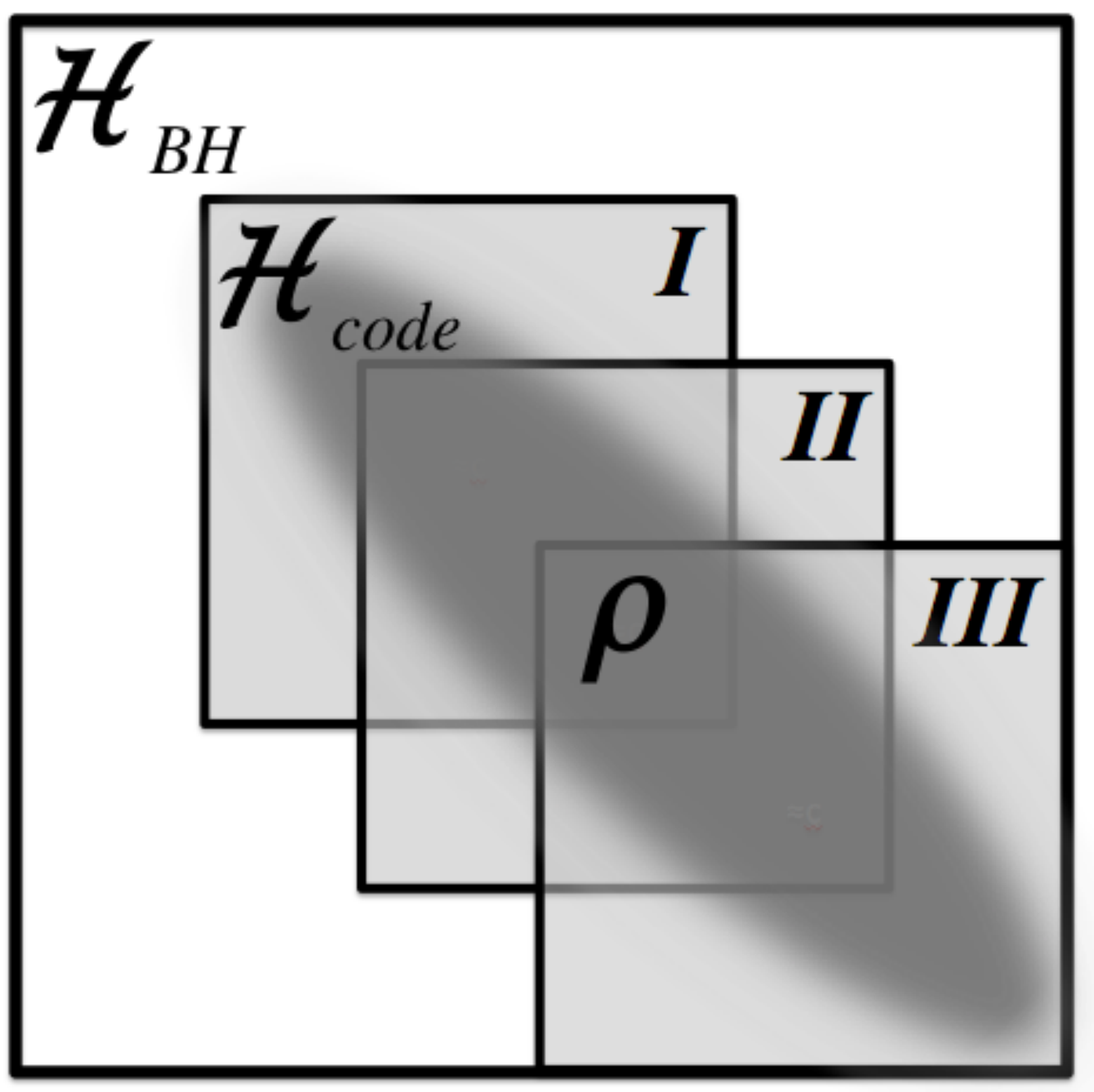}
\end{center}
\begin{center}
\caption{\small
{
The QECC protects the coherence of a subspace ${\cal H}_{\rm code}$ within the black hole Hilbert space ${\cal H}_{BH}$. It safeguards the smoothness of the horizon for every state in ${\cal H}_{\rm code}$. As the density matrix 
$\rho$ spreads out, the code subspace  needs to grow along with it. This degrades the fidelity of the code and spoils the semi-classical correspondence. The state decoheres into a sum of semi-classical components, each of which fits inside a smaller code space and has a smooth horizon.}}
\end{center}
\vspace{-0.5cm}
\end{figure} 

However, we have also found that when the von Neumann entropy of the internal black hole state becomes too large, then the quantum recovery 
operation receives an error contribution of the form (\ref{newerror}). The low energy effective field theory breaks down.
This is consistent with the firewall argument. But we have learned more: with our methods we have been able to qualify and quantify how
 the effective field theory description begins to break down as the entanglement with the external environment increases.

What we have just described is a standard issue in quantum measurement theory. Suppose we think of the black hole as a measuring device.  
We can then  ask under which circumstance can it measure the state of the radiation. 
Measurement in general happens when the measured object decoheres by entangling itself with the internal quantum states of the measuring device. 
The firewall argument invokes the situation where the complete state of a big measuring apparatus $A$ (black hole) itself has been measured
in the past by an even bigger system $Z$ (early radiation). Monogamy of entanglement then seems to imply that the smaller apparatus $A$ can 
no longer do  any measurements on an even smaller system $B$ (late radiation). This is not true, in general.

It is useful to compare two situations: (a) the bigger system $Z$ remains in sufficient contact with the smaller apparatus
$A$, so that it keeps actively measuring the complete state of $A$,
or (b) $A$ was once maximally entangled with $Z$ in the far past, but $A$ and $Z$ are dynamically decoupled at present.
In  case (a),  $A$ can indeed not do any measurements itself. This follows from the dynamics of dephasing: in other
to measure some other small system, $B$, it needs to build phase coherence with $B$. But while $A$ is being measured, all components of 
its state are randomized in ways that prevent its ability to build phase coherence with anything else. So in this case, entanglement is indeed
monogamous. In case (b), however, the measuring apparatus $A$ can in general still do measurements. Once it is dynamically decoupled 
from $Z$, it can start building  new phase coherence with some other system $B$. So  even if $A$ started out in a maximally mixed state, 
over time it can build new entanglement via interactions with its environmen, and measure $B$. The black hole situation is more like case (b).

\section*{Acknowledgements}

We thank  Bartek Czech,  Borun Chowdhury, Jan de Boer, Daniel Harlow, Steve Jackson, Juan Maldacena and Andrea Puhm for
helpful discussions.   The work of E.V. is supported by a Spinoza Grant of the Dutch Science Foundation (NWO), an Advanced Grant by the European Research Council (ERC) and the Foundation for Fundamental Research on Matter (FOM). The work of H.V. is supported by NSF grant PHY-0756966.

\bigskip

\appendix

\section*{Appendix A: Some calculations}

In this Appendix we outline some of the calculations, the were left out of the main text in Chapter 3. We first do the calculation of the expectation value (\ref{unruhvev})
of the product $\bfA B$ of an internal operator $\bfA$ given in (\ref{adef}) and an external operator $B$ made out of $b$-modes. We first do an intermediate
calculation 
\bea
{}_\bbbb \la  0 \spc   \bigr| {\textit{\textbf U}}^{\, \dag}  \textit{\textbf A} B \spc {\textit{\textbf U}} \, \bigl| 0 \ra_{\bbbb}   \is {}_\bbbb \la  0 \spc   \bigr| {\textit{\textbf U}}^{\, \dag} \,  {}_\aaaa\la  0  \bigr|\rR^\dag  A\spc \spc \rR \smpc \bigl|  0 \ra_{\aaaa}  \; B \spc {\textit{\textbf U}} \, \bigl| 0 \ra_{\bbbb}   \\[2mm]
 \is
{}_\bbbb\la  0  \bigr| {}_\aaaa\!\smpc\la  0  \bigl|\spc \sS^\dag\rR^\dag \spc A\spc B\spc \rR \sS\spc \bigl|  0  \ra_{\aaaa} \bigl| 0\ra_{\bbbb}   \\[2mm]
\is \la \spc 0_{\spc U}\spc \bigr| A B \bigl|\, 0_{\spc U}\ra \; \mathbb{1}_{\rm code}.  
\eea
Here we first use the definition (\ref{adef}), then move the B operators inwards (since they commute with the recovery operators), and finally use the
result (\ref{recoveryr}) for the action of $\rR\uU$ on states of the form $\bigl|\spc i \spc \ra\spc \bigl| 0 \ra_{\aaaa} \bigl| 0\ra_{\bbbb}$. Here $\mathbb{1}_{\rm code}$ denotes
the unit operator on ${\cal H}_{\rm code}$. Now we take the trace with the density matrix $\rho(0)$ given in (\ref{rhosmall}). Since $\rho(0)$ is assumed to
act within ${\cal H}_{\rm code}$, we immediately find
\bea
 \; \tr\bigr(\rho(\tau) \spc  \textit{\textbf A} B \bigr) 
 \is   \sum_{\ibar,\kbar} \rho_{\ibar\! \kbar} \la \ibar \bigr|  \la  0_\bbbb \spc   \bigr| {\textit{\textbf U}}^{\, \dag}  \textit{\textbf A} B \spc {\textit{\textbf U}} \, \bigl| 0_{\bbbb}  \ra \bigl|\kbar\ra \\[2mm]
 \is   \sum_{\ibar, \kbar} \rho_{\ibar\! \kbar}\spc \la \ibar \bigr|\la \spc 0_{\spc U}\spc \bigr| A B \, 0_{\spc U}\ra|\kbar \ra  = \la \spc 0_{\spc U}\spc \bigr| A B \, 0_{\spc U}\ra.
\eea
This reproduces the expectation value in the local Minkowski vacuum.

Next we will verify eqn (\ref{pialgebra}). We will work in the limit that ${N_{\rm code}}\ll N$. Using the definition (\ref{recover}), we have 
\be
R_{\nom}^\dag R_\mom =  \frac 1 {\sqrt{\pp_\nomt\pp_\momt}} \sum_{\ibar} \cC_\nom\spc  \bigl|{} \ibar{} \ra \la{} \ibar{} \bigl|  \cC^{\, \dag}_\mom.
\ee
We now compute
\bea
R^\dag_\nom R_\som\, R^\dag_\tom R_\mom \spc \spc \is    \frac 1 {\sqrt{\pp_\nom \pp_\som \pp_\tom \pp_\mom} }
\sum_{\ibar, \kbar} \cC_\nom{}  \bigl|{} \ibar{} \ra \la{} \ibar{} \bigl| \overlinde{ \cC^{\, \dag}_\som  \cC_\tom}\spc  \bigl|\kbar \ra \la \kbar \bigl|  \cC^{\, \dag}_\mom \\[3mm]
\is \frac {\delta_{\som\tom}} {\sqrt{\pp_\nom\pp_\mom}} \sum_{\ibar} \cC_\nom\spc  \bigl|{} \ibar{} \ra \la{} \ibar{} \bigl|  \cC^{\, \dag}_\mom \; = \; 
  \delta_{\som\tom} \, R^\dag_\nom R_\mom,
\eea
where we used eqn (\ref{keyrel}). In the same way, we derive (\ref{rhobb})
\bea
 \tr_\aaA\! \bigl(\rho \, R^\dag_\nom R_\mom \bigr)  \is  \sum_{\ibar,\kbar,
\bar{s}} \rho_{\ibar\! \kbar} \frac{1}{\sqrt{\pp_\tom\pp_\som}} \la \spc \ibar \spc \bigl| \overlinde{\cC^{\, \dag}_\mom \cC_\som}\spc  \bigl| \bar{s} \ra \la\bar{s} \bigl| \overlinde{ \cC^{\, \dag}_\tom \cC_\nom}\spc \bigl| \, \kbar \, 
\ra \\[3mm]
\is   \sqrt{\pp_\nom\pp_\mom} \, \delta_{\nom\tom}
\, \delta_{\mom\som},
\eea
where we use (\ref{keyrel}) twice.

%%%%%%%%%%%%%%%%%%%%%%%%%%%%%%%%%%%%%%%%%%%%%%%%

%%%%%%%%%%%%%%%%%%%%%%%%%%%%%%%%%%%%%%%%%%%%%%%%
\end{document}